# Energetic Particle Acceleration and Transport: Interplanetary Coronal Mass Ejections and Shocks

OLGA E. MALANDRAKI[1]*, CHRISTINA M. S. COHEN[2], JOE GIACALONE[3], LENG YING KHOO[4], DOMENICO TROTTA[5], LAURA RODRIGUEZ-GARCIA[5,6], ATHANASIOS KOULOUMVAKOS[7], ZHEYI DING[8], YUNCONG LI[9,10], ALEXANDER KOLLHOFF[8], JOHN G. MITCHELL[11], DAVID J. MCCOMAS[4], NATHAN A. SCHWADRON[12,13], JAVIER RODRIGUEZ-PACHECO[6], ROBERT F. WIMMER-SCHWEINGRUBER[8], GEORGE C. HO[14], GLENN MASON[7], MICHALIS KARAVOLOS[1], and JINGNAN GUO[9,15]

[1]National Observatory of Athens, IAASARS, Athens, Greece.
[2]California Institute of Technology, Pasadena, United States.
[3]University of Arizona, Tucson, United States
[4]Department of Astrophysical Sciences, Princeton University, Princeton, United States
[5]European Space Agency (ESA), European Space Astronomy Centre (ESAC), Villanueva de la Cañada, Madrid, Spain
[6]Universidad de Alcalá, Space Research Group (SRG-UAH), Alcalá de Henares, Madrid, Spain
[7]The Johns Hopkins University, Applied Physics Laboratory, Maryland, Laurel, United States
[8]Institut für Experimentelle und Angewandte Physik, Christian-Albrechts-Universität zu Kiel, Kiel, Germany
[9]National Key Laboratory of Deep Space Exploration/School of Earth and Space Sciences, University of Technology and Science of China, Hefei, China
[10]Institute of Physics, University of Graz, Graz, Austria
[11]NASA, Goddard Space Flight Center, Greenbelt, MD 20771, United States
[12] Los Alamos National Laboratory, PO Box 1663, Los Alamos, NM 87545, USA
[13] University of New Hampshire, Durham, United States
[14]Southwest Research Institute, San Antonio, United States
[15]CAS Center for Excellence in Comparative Planetology, University of Science and Technology of China, Hefei, China
*Corresponding author. E-mail: omaland@noa.gr

**Abstract.** Solar Energetic Particles from suprathermal (few keV) up to relativistic (few GeV) energies constitute an important contributor to the characterization of the space environment. Emitted from the Sun they are associated with solar flares and shock waves driven by Coronal Mass Ejections. This review presents important recent results of the study of Interplanetary CMEs and shocks in relation to energetic particle acceleration and transport, taking advantage of multi-point, multi-instrument observations available by a fleet of spacecraft in the heliosphere. In particular, the Solar Orbiter and Parker Solar Probe pioneering missions providing unprecedented measurements of energetic particles in the near-Sun environment offer promising insights into several unresolved questions particularly regarding the mechanisms responsible for the acceleration and transport of energetic particles in the heliosphere and their relative contributions.

## 1. Introduction

Solar Energetic Particles (SEPs) from suprathermal (few keV) up to relativistic (few GeV) energies are sporadic enhancements of particle intensities associated with transient solar activity (Reames, 1999; Desai & Giacalone, 2016; Cohen et al. 2021; Zhang et al. 2021). They provide critical insights into particle acceleration and transport mechanisms in the heliosphere. Large SEP events are typically associated with coronal mass ejection (CME)-driven shock waves. These shocks are efficient sites for particle acceleration. A widely accepted mechanism for proton acceleration during large SEP events is diffusive shock acceleration (DSA) (Axford et al. 1977; Drury 1983). In the DSA mechanism, particles gain energy by crossing the shock front multiple times, with the maximum particle energy being controlled by the scattering efficiency and the finite acceleration time. Following their escape from the traveling shock front, particles propagate along the Interplanetary Magnetic Field (IMF) undergoing a series of transport processes, such as pitch-angle scattering, adiabatic momentum changes, and magnetic focusing/mirroring, which shape their spatio-temporal and energy distributions (Li et al. 2003; Wijsen et al. 2019; Hu et al. 2018). Analysis of widespread SEP events in the heliosphere has shown that the spread of SEPs over a wide range of heliolongitudes can be due to a combination of different processes, including the extent of the shock wave. Furthermore, particles may also diffuse perpendicular to the IMF, enabling cross-field transport and allowing particles to spread widely in longitude and latitude (e.g. Dwyer et al. 1997; Strauss et al. 2017; Wang et al. 2012). Understanding the mechanisms that accelerate and transport SEPs in the heliosphere is also an important element of space weather forecasting to ensure the success of future space exploration by humans (Whitman et al. 2023; Guo et al. 2024).

In this work, recent important advances and results in the study of the energetic particle acceleration and transport in relation to CME-driven shocks are reviewed by means of multi-instrument observations by a fleet of spacecraft in the heliosphere including Solar Orbiter, Parker Solar Probe, MAVEN (Larson et al. 2015), STEREO (Luhmann et al. 2008; Howard et al. 2008), SOHO (Brueckner et al. 1995), Wind (Ogilvie and Desch, 1997), ACE (Stone et al. 1998), BepiColombo (Pinto et al. 2022; Heyner et al. 2021). In particular, relevant results with the unprecedented measurements provided by the Parker Solar Probe (PSP; Fox et al. 2016) and Solar Orbiter (SolO; Mueller et al. 2020) spacecraft well inside 1 AU are also reviewed. It is noteworthy that the PSP and SolO missions have opened a novel observational window for Interplanetary (IP) shocks at previously unexplored, small heliocentric distances. The Integrated Science Investigation of the Sun (IS⊙IS; McComas et al. 2016) is the energetic particle suite onboard PSP, consisting of two main energetic particle instruments (EPIs), EPI-Lo (Hill et al. 2017) and EPI-Hi (Wiedenbeck et al. 2017). At PSP, magnetic field data are used from the FIELDS instrument (Bale et al. 2016), while proton density, bulk flow velocity, and temperature are obtained by the Solar Wind Electrons Alphas and Protons (SWEAP) investigation suite (Kasper et al. 2016). At SolO the Energetic Particle Detector (EPD; Rodríguez-Pacheco et al. 2020) is used to investigate the properties of energetic particles at SolO, the magnetic field is measured by the fluxgate magnetometer (MAG; Horbury et al. 2020), the ion bulk flow speed, plasma density and temperature and electron distribution functions are obtained from the Solar Wind Analyser suite (SWA; Owen et al. 2020).

In the following sections, we review the main results on shock properties and SEPs in the inner heliosphere during the 2022 September 5 SEP event, from both the observational and modeling perspective, the unique multi-spacecraft observations of the widespread SEP event on 2022 February 15-16, and the 2022 January 20 widespread SEP event in which SEPs are injected inside and outside a magnetic cloud. Particular emphasis is then placed on the critical high-resolution observations of SEP events in the inner heliosphere exhibiting an 'inverse velocity arrival' or 'inverse velocity dispersion' feature, as well as detailed discussion of the underlying mechanisms of particle acceleration and transport that lead to this unusual phenomenon.

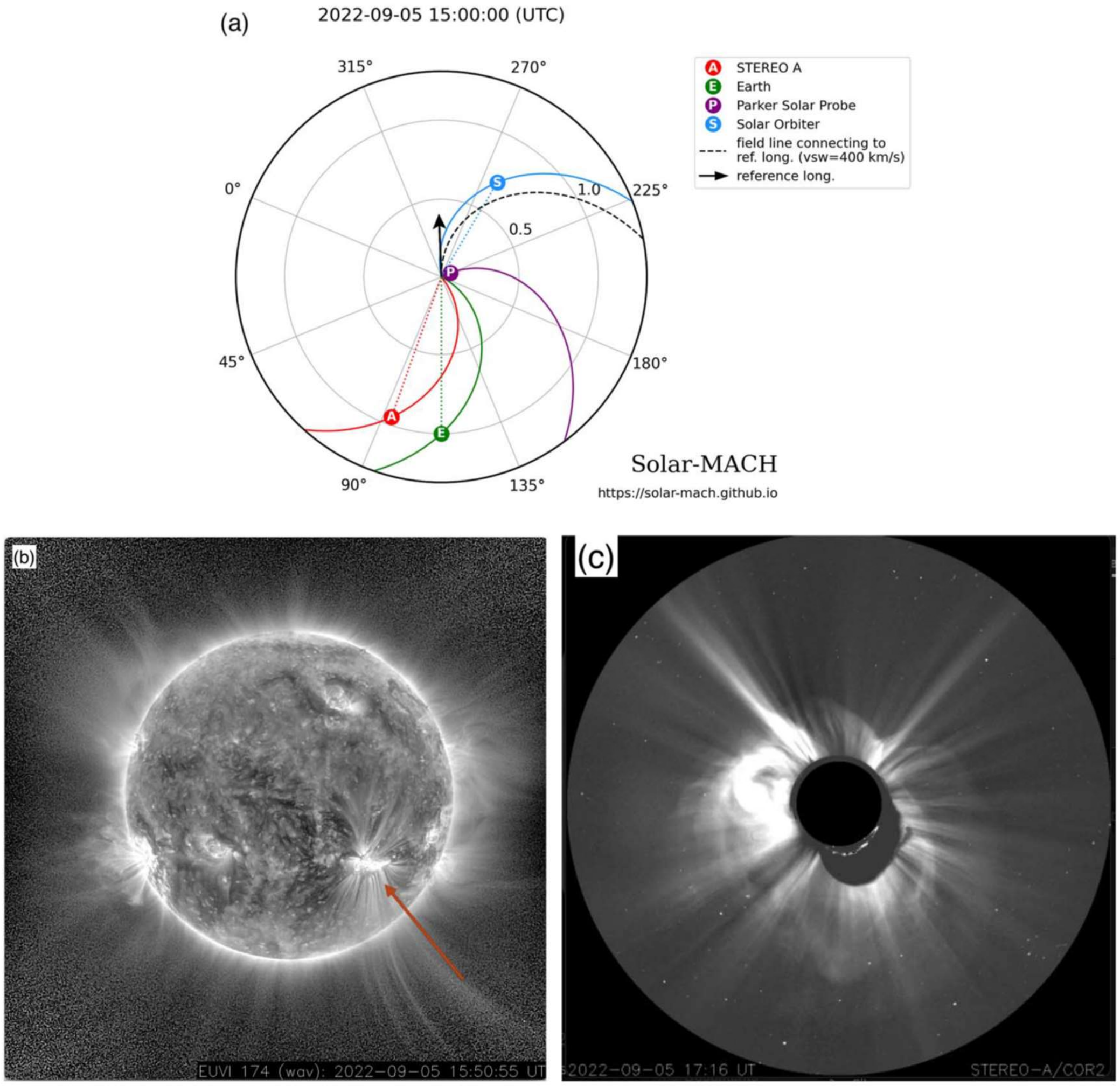

**Figure 1**. (a) Spacecraft configuration at 15:00 UT on 2022 September 5 in the Carrington coordinate system using the Solar-MACH tool (Gieseler et al. 2023). The Sun is at the center and the Earth is represented by the green circle (E). The following spacecraft positions are also plotted: STEREO-A (A, red), Parker Solar Probe (P, purple), and Solar Orbiter (S, blue). The black arrow indicates the longitude of the parent active region and the colored lines (and dashed black line for the active region) represent Parker spiral interplanetary magnetic field lines for a nominal 400 km $s^{-1}$ solar wind connecting each spacecraft with the Sun. (b) EUV image of the Sun from SolO/EUI; the parent active region is indicated by the red arrow. (c) Difference image of the corresponding CME as observed by SECCHI/COR2 onboard STEREO-A. Adapted from Cohen et al. (2024).

## 2. Shock properties and SEPs in the inner heliosphere – 2022 September 5 event

The 2022 September 5 CME event provided an opportunity to study a CME-driven IP shock at unprecedentedly low heliocentric distance, with the CME-driven IP shock crossing PSP when at 0.07 AU. The related SEP event exhibited unique features challenging our current understanding, based on the 'inverse velocity arrival' feature observed. SolO, at 0.7 AU, being radially well-aligned with PSP (Figure 1(a)) provided important insights about the evolution of the shock and the shock-accelerated proton characteristics.

An extremely fast and wide CME associated with Active Region AR13102 erupted from the Sun and into the IP space on 2022 September 5 (Cohen et al. 2024). The CME was not Earth directed, as the eruption happened on the far side of the Sun. An overview of the spacecraft configuration in the inner heliosphere is shown in Figure 1(a). Solar Orbiter was well-positioned to view the AR and the flare (Figure 1(b)) estimated as a GOES class equivalent of M6-X2 (Cohen et al. 2024). The CME was observed as a backside halo by LASCO onboard SOHO and STEREO-A/SECCHI (Figure 1(c)). Figure 2 of Kouloumvakos et al. (2025) shows EUV and white-light observations, with an image at 304 Å from SolO/EUI/FSI at 16:15 UT on 2022 September 5, in which the flare ribbons are visible at the bottom right of the EUI image, where the AR was located, as well as the well-developed CME and shock wave in coronagraphic images from LASCO/C2 and STEREO-COR2 at 16:48 UT and 16:56 UT on 2022 September 5, respectively. As Paouris et al. (2023) and Cohen et al. (2024) reported, from the STEREO-COR2 coronagraph observations, the speed of the CME was determined to be > 2200 km s$^{-1}$, placing it in the top 1% of the speed distribution for all CMEs observed by LASCO from 1996 January 22 to 2022 July 1 (list compiled by A. Vourlidas, private communication).

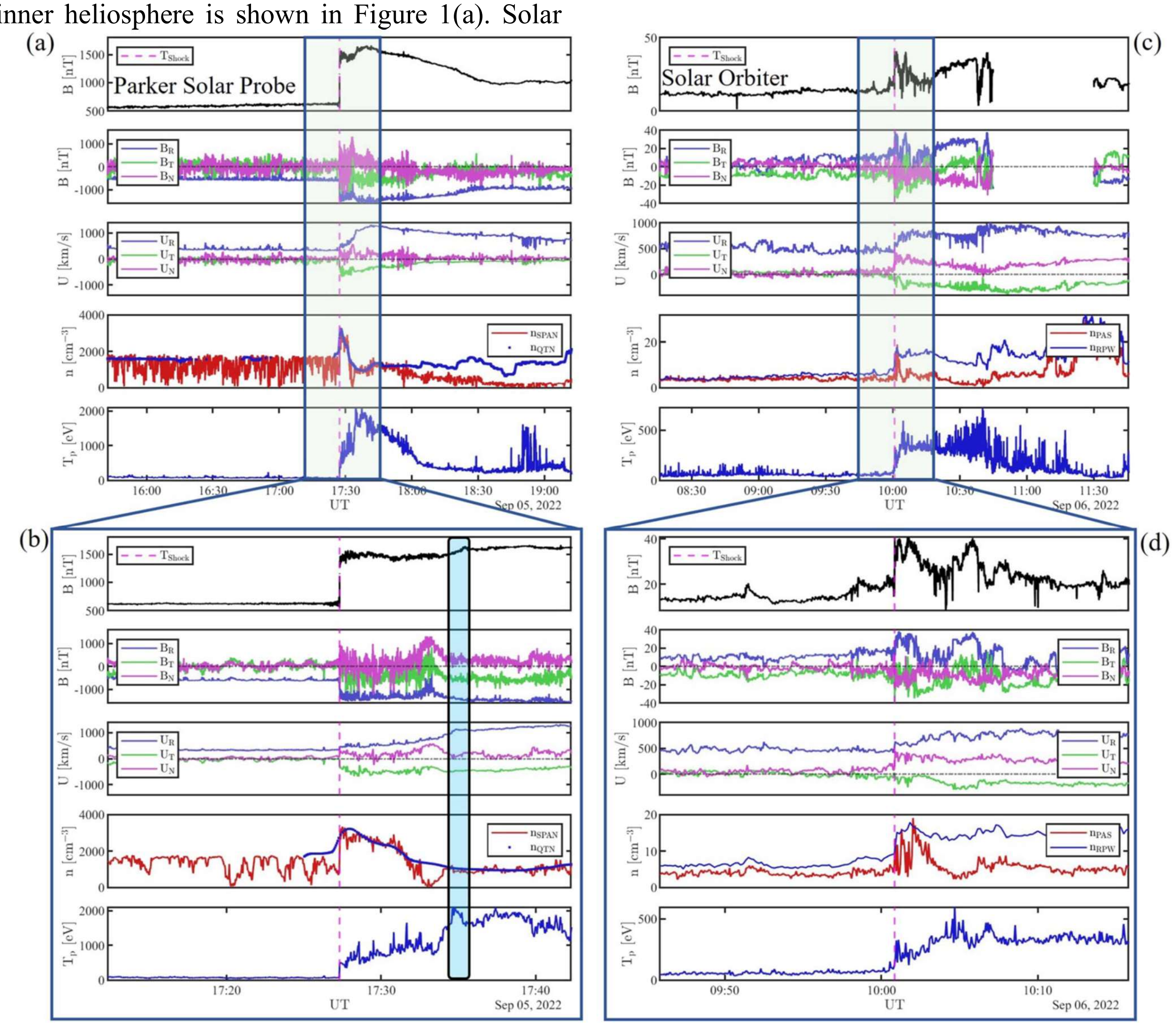

**Figure 2.** Magnetic field magnitude and components in RTN, ion bulk flow velocity, density and temperature as measured by PSP (left) and SolO (right). An overview of the event (a, c) is shown as well as a zoom-in around the shock transition for both spacecraft, marked by the dashed magenta line. The blue shaded area in panel (b) indicates the approximate end of the CME sheath as observed by PSP (see text). Reproduced from Trotta et al. (2024).

2.1 *The shock at Parker Solar Probe and Solar Orbiter*

A fast forward shock, driven by the CME, was detected by PSP at an unprecedentedly low heliocentric distance of 0.07 AU (~15 $R_\odot$) at 17:27:19 UT on 2022 September 5. An overview of the in situ magnetic field and plasma quantities as measured by PSP (left) and SolO (right), as well as a zoom-in around the shock transition at both spacecraft are shown in Figure 2 (Trotta et al. 2024). At PSP, a sharp transition in magnetic field, with a jump from ~600 to 1600 nT, occurs at the shock. The solar wind velocity profile exhibits a large deflection in the transverse direction $V_T$, observed in correspondence with the shock crossing, and a steady rise in the radial component $V_R$ further downstream, up to ~17:35 UT that has been marked as the end of the CME sheath region. However, it is noted that the boundary between the sheath and CME flux rope is not clear for this event, with the blue shaded area in panel (b) of Figure 2 indicating the approximate end of the CME sheath region. The shock at PSP has a notably short sheath region (~ 5 min) with respect to typical sheaths at 1 AU (lasting hours see e.g. Ala-Lahti et al. 2020), due to its early evolutionary stage, an interesting property that may have fundamental implications for the possibility of accelerating particles to high energies (see section 2.2).

In Situ Shock Parameters at PSP and SolO

| | Parker Solar Probe | Solar Orbiter |
|---|---|---|
| Shock Time (UT) | 2022 September 5, 17:27:19 | 2022 September 6, 10:00:51 |
| Carrington longitude (deg) | 232 | 251.9 |
| Carrington latitude (deg) | −1.8 | −3.6 |
| Heliocentric dist. (au) | 0.07 | 0.7 |
| $\langle \hat{n}_{RTN} \rangle$ | [0.5 −0.8 0.2] | [0.6 −0.2 0.7] |
| $\langle \theta_{Bn} \rangle$ (deg) | 53 | 51 |
| $\langle r_B \rangle$ | 2.3 | 1.9 |
| $\langle r \rangle$ | 1.6 | 2 |
| $\langle v_{sh} \rangle$ (km s$^{-1}$) | 1520 | 942 |
| $\beta_{up}$ | 0.1 | 0.6 |
| $M_{fms}$ | 3.8 | 3.2 |
| $M_A$ | 3.9 | 3.8 |

**Table 1.** From top to bottom: shock arrival time, Carrington longitude and latitude and heliocentric distance at the time of shock arrival, shock-normal vector in RTN coordinates, shock-normal angle $\theta_{Bn}$, magnetic compression ratio $r_B$, gas compression ratio $r$, shock speed $v_{sh}$, upstream thermal to magnetic pressure ratio plasma beta $\beta_{up}$, and the fast magnetosonic and Alfvénic Mach numbers ($M_{fms}$ and $M_A$), respectively. Reproduced from Trotta et al. (2024).

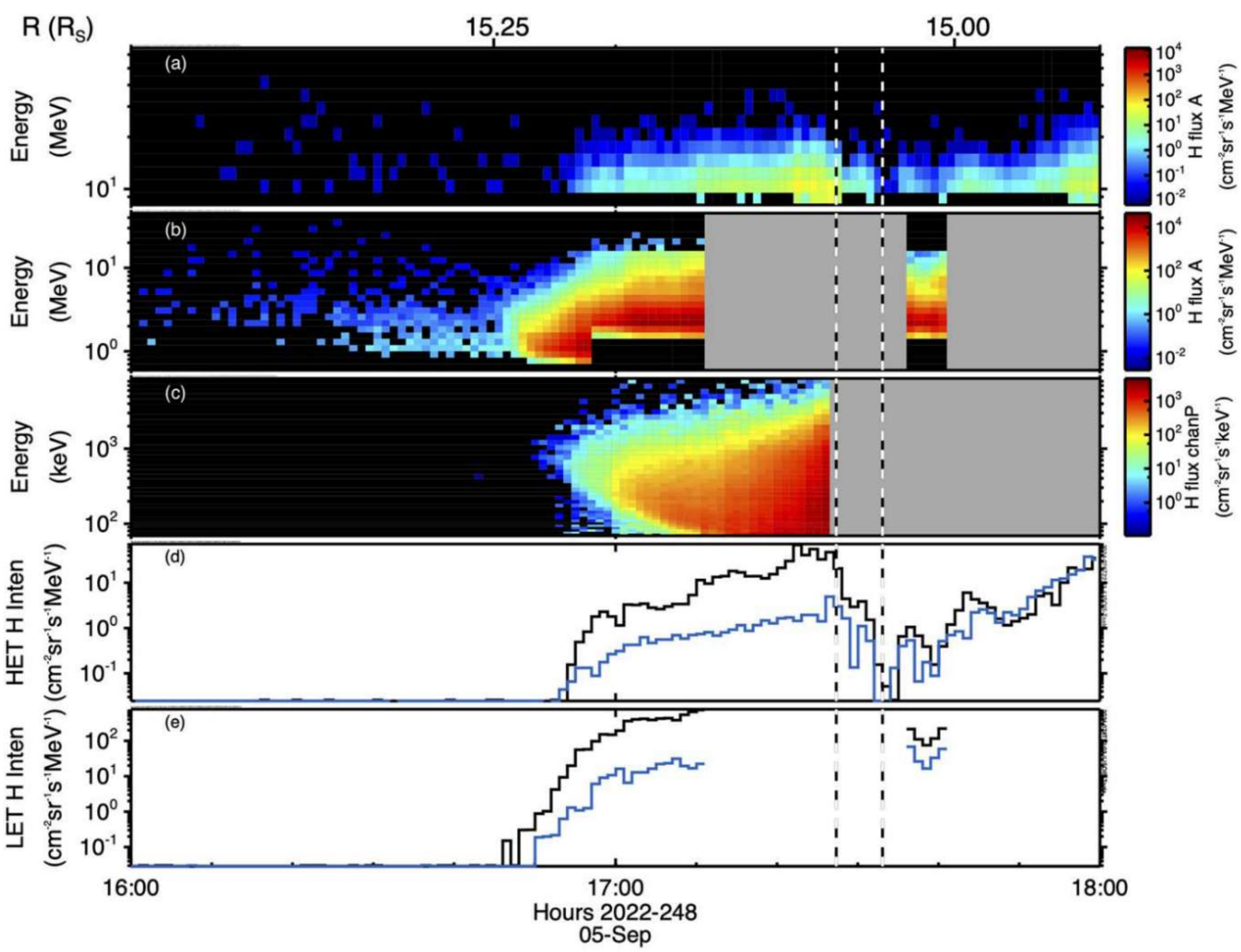

**Figure 3.** Energy spectrograms of proton intensities from the IS⊙IS instrument onboard PSP. The spacecraft distance from the Sun is shown along the x-axis on top in solar radii. (a) HETA, (b) LETA, and (c) EPI-Lo. Proton intensities for (d) 9.5-11.3 MeV from HETA (black) and HETB (blue) and (e) for 4.8-5.7 MeV from LETA (black) and LETB (blue). Shock arrival at 17:27 UT and the end of the sheath at 17:33 UT are indicated by dashed line. (Trotta et al. 2024, however, state that the boundary between the sheath and the start of the flux rope is not very clear, see text in section 2.1). Reproduced from Cohen et al. (2024).

A comprehensive characterization of the shock parameters locally observed by PSP was carried out by Trotta et al. (2024) (Table 1). The estimated average shock normal $\langle n \rangle$ of the ensemble of shock normals computed by varying the upstream / downstream averaging windows from a few seconds to a few minutes (see Trotta et al. 2022) significantly departing from the radial direction, indicates that PSP likely crossed a flank of the shock. The estimation of $\theta_{Bn}$ i.e. the angle between the shock normal and the upstream magnetic field, a crucial parameter influencing shock acceleration efficiency revealed the presence of an oblique ($\theta_{Bn}$ ~ 53°) supercritical shock (e.g. Burgess and Scholer, 2015) with moderate Alfvénic and fast magneto-sonic Mach numbers ($M_A \approx 3.9$, $M_{fms}$ ~ 3.8, respectively) with respect to other IP shocks observed at or near 1 AU. Furthermore, it was found that the shock propagates at the very high speed of ~1500 km $s^{-1}$ in the spacecraft frame and along the estimated shock normal. Using PSP observations, Trotta et al. (2024) also investigated how magnetic switchbacks and ion cyclotron waves are processed across shock crossing, a topic outside the scope of this review.

The shock crossed SolO at 10:00:51 UT on 2022 September 6. As shown in Figure 2 right, the shock crossing at SolO appears much more structured compared to the PSP observations. As shown by Trotta et al. (2024) steep upstream enhancements of magnetic field magnitude are observed ahead of the shock (09:52, 09:56 UT) compatible with very rarely observed shocklets at strong IP shocks. A high degree of magnetic field structuring has also been found downstream of the shock, indicating a

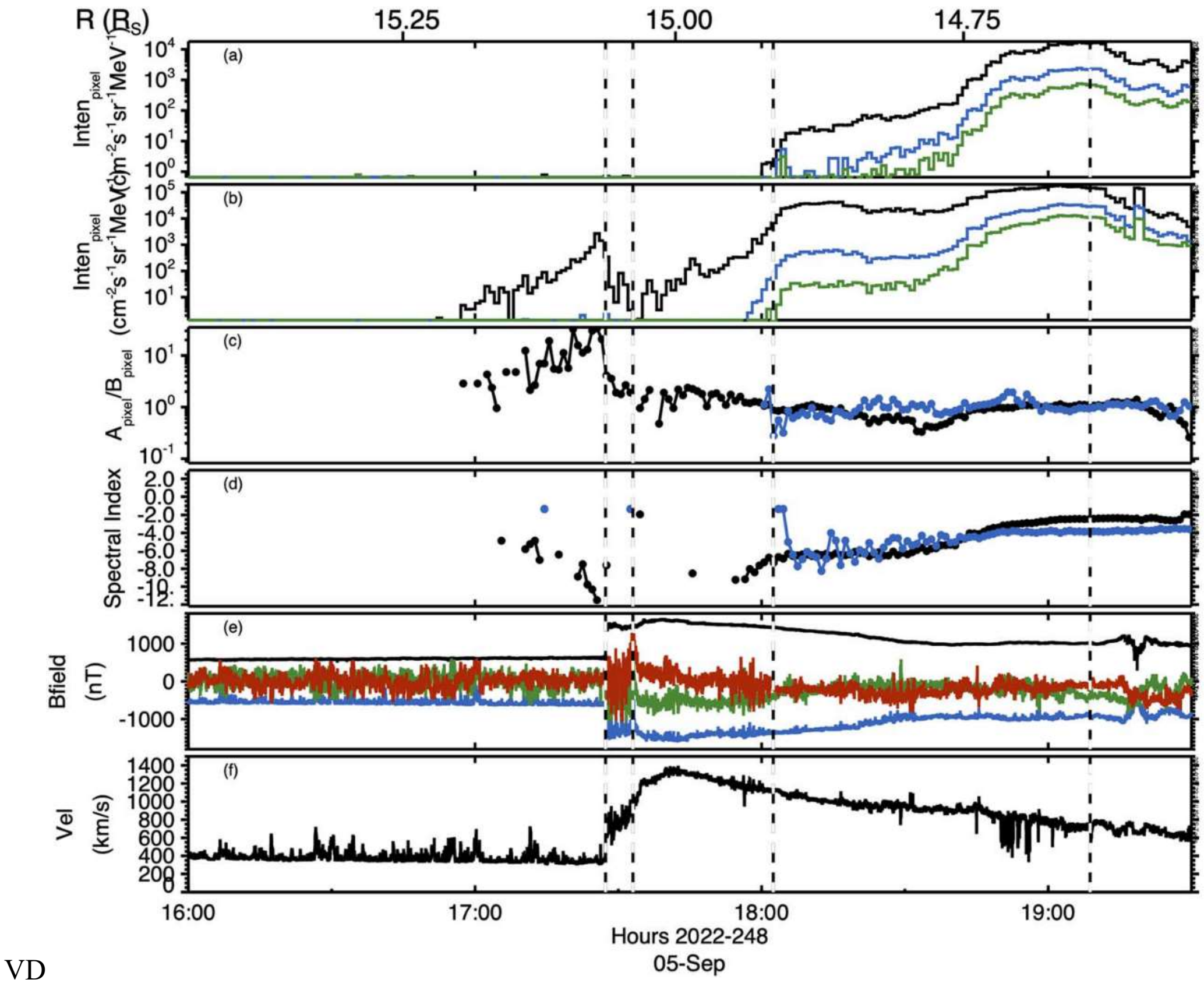

VD

**Figure 4.** Evolution of ion intensities by the PSP spacecraft in pixel rates from (a) HETA at ~15 MeV (black), ~25 MeV (blue) and ~33 MeV (green); (b) LETA at ~8 MeV (black), ~15 MeV (blue), and ~21 MeV (green). Changes in anisotropy shown in (c) ratios of A-side pixels for LET at ~8 MeV (black) and HET at ~15 MeV (blue). Changes in spectral index shown in (d) ratios of LETA (black) and HETA (blue) deeper pixels to shallower pixels; (e) Magnetic field magnitude (black) and components in RTN; (f) solar wind velocity. Dashed lines indicate (from left to right) the times of the shock arrival, the end of the CME sheath, the crossing of the CME flank, and the entry into a region of closed field lines internal to the CME identified as a magnetic cloud. The PSP distance from the Sun in solar radii is given along the top x-axis. Reproduced from Cohen et al. (2024).

high level of complexity for this shock crossing. The shock environment at SolO is much more disturbed than the one observed at PSP. A shock parameter estimation using ~1 minute averaging windows, thus addressing the very local shock properties, yields similar values to the PSP parameters (Table 1). However, analyzing the 30 minutes across the shock transition parameters are found compatible with a quasi-parallel shock transition ($\theta_{Bn}$ ~ 30°) with high Mach numbers ($M_A \sim M_{fms} \sim 7$). In addition to the local variability examined, it is to be expected that the characteristics of the shock will differ at the points where it crosses two separated spacecraft such as PSP and SolO separated by ~20° in longitude and by a large radial distance (0.63 au).

### 2.2 *Particle acceleration at Parker Solar Probe and Solar Orbiter*

Previous observations have shown that if the accelerated particles are released nearly simultaneously, high-energy particles arrive at the observer earlier than lower-energy particles. This behaviour is known as Velocity Dispersion (VD) and is often observed in the initial phase of SEP events (e.g. Malandraki et al. 2012; Gómez-Herrero et al. 2021; Wimmer-Schweingruber et al. 2023). Velocity Dispersion Analysis (VDA) assumes the scatter-free propagation of particles along magnetic field lines, allowing their energy-dependent arrival times to be used for estimating the particle release time and the effective path length (Tylka et al. 2003; Laitinen et al. 2015).

The SEP event on September 5, 2022, offered a compelling case for understanding SEP acceleration and transport close to the Sun. Cohen et al. (2024) presents an overview of the PSP observations of this event which exhibited several unexpected characteristics. As shown in Figure 3, the onset of the SEP event occurs around 16:45 UT on 2022 September 5. A notable feature is the shape in energy versus time. The onset of the event was found to have normal velocity dispersion at energies below ~1 MeV (panel (c)), however above that energy the higher energy particles arrived later than their lower energy counterparts, evident in all three panels (a) - (c), producing a feature termed 'inverse velocity arrival'. Cohen et al. (2024) have argued that the dispersion shape of the SEP onset is well explained by a diffusive shock acceleration process. It requires time for particles to be accelerated to high energies. Therefore, if an observer is sufficiently close to a shock undergoing particle acceleration, lower-energy particles, which have undergone a shorter acceleration process, are likely to reach the observer first, followed by higher-energy particles. The balance between the timescale for acceleration and the timescale for transit from the shock to the spacecraft would determine the energy at which the 'inverse velocity arrival' feature starts to be observed (in this case ~1 MeV). A number of recent studies have performed shock and SEP modeling for this event (e.g., Kouloumvakos et al. 2025; Do et al. 2025) (see more details below) and the cause of this feature appears to be most likely due to the inefficiency of the shock to accelerate high energy particles early in its development (i.e., close to the Sun where it was observed by PSP).

When the shock crossed PSP, observations of SEPs below ~8 MeV were not available due to instrumental configurations. However, the intensities of SEPs with higher energies all plunged by orders of magnitude coincident with the shock passage as shown in Figure 4(b). Figure 3(a) shows a similar pattern of increasing intensities as the shock approaches and a substantial drop at its passage. Such a dramatic decrease was unexpected and remains to be explained. The intensities only began to recover after the spacecraft exited the CME sheath region and crossed the flank of the CME, peaking when PSP entered the region of closed magnetic field lines internal to the CME as shown in Figure 4.

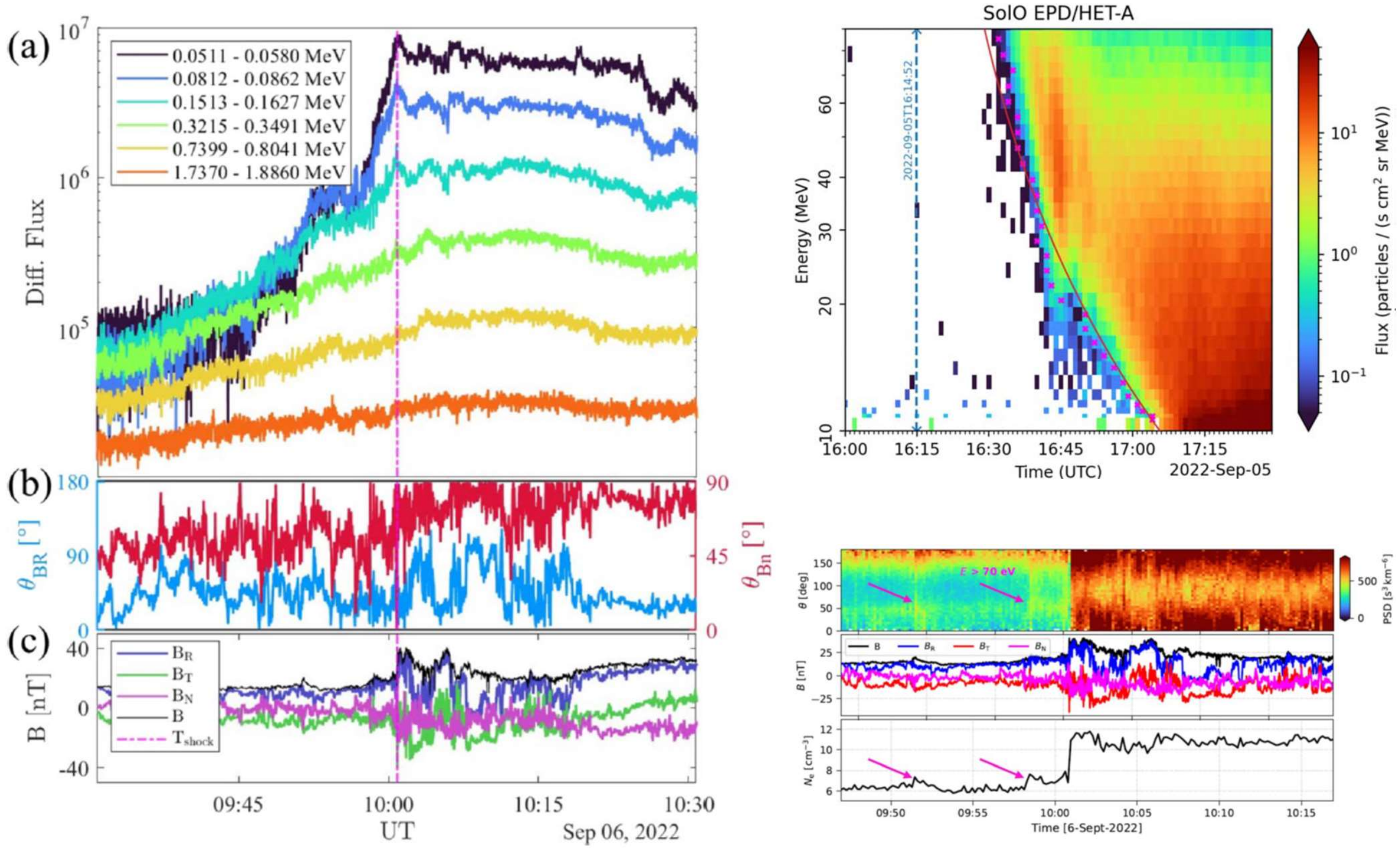

**Figure 5.** Left: (a) ~0.05-2 MeV energetic ion fluxes as observed by the SolO/EPD-EPT instrument. (b) local $\theta_{BR}$ (blue) and $\theta_{Bn}$ (red) angles. (c) magnetic field magnitude (black) and its components. Right top panel: Energy spectrogram of ~10 to ~90 MeV proton fluxes from SolO/HET-Sun telescope. The red line is the VD curve fitted to the SEP onset times depicted with the magenta crosses. The vertical dashed line denotes the SEP release time from the VDA, shifted from the Sun to the spacecraft taking into account the travel time to each observer. Right bottom panel: Top: SolO/EAS PADs for >70 eV particles, Middle: Magnetic field measurements in RTN, Bottom: Electron density computed from SolO/EAS measurements. Adapted from Trotta et al. (2024) and Kouloumvakos et al. (2025).

Finally, as Cohen et al. (2024) reported, there were several periods within the CME in which PSP observed significant anisotropies in the SEPs, with flow in the sunward direction dominating flow in the anti-sunward direction by factors of several. These periods lasted for ~10 minutes and were not clearly associated with any feature observed in the magnetic field or solar wind plasma. Currently these features remain unexplained, but future analysis may be able to assess whether these sunward flows were spatially contained (e.g. within a given flux tube that PSP moves through) or a result of an intermittent connection to an acceleration region beyond the spacecraft (possibly the shock or some turbulence/compressions found within the CME).

SolO observed an intensity increase at shock passage up to 2 MeV. Figure 5a (left) shows energetic ion fluxes in the ~0.05-2 MeV energy range as measured by the Sunward telescope of the EPD/EPT instrument onboard SolO as well as magnetic field parameters during the time of the shock crossing (indicated by the dashed magenta line). Figure 5b (left) shows the $\theta_{Bn}$ angle (red), i.e. the angle between the shock normal and the local magnetic field, as well as the local $\theta_{BR}$ angle (blue) which is the angle between the radial direction and the local magnetic field. Structuring observed in the shock surroundings is particularly evident in the 20 minutes downstream of the shock, populated with many sharp changes of the magnetic field. SolO observed energetic ions of up to ~2 MeV accelerated by the shock with fluxes rising as the shock crossing is approached, a typical behaviour of ion acceleration at IP shocks (e.g. Giacalone, 2012). Some of the fluxes (e.g. ~0.33 and ~0.7 MeV shown in green and yellow time intensity profiles respectively) exhibit their peaks at times not coincident with the shock passage at SolO, but instead downstream of the shock. This behaviour may well be due to additional acceleration mechanisms operating downstream of the shock (Zank et al. 2015a; Zhao et al. 2018). A crucial feature is the fluctuations found in the fluxes of the energetic particle population immediately downstream of the shock, with typical timescales of 1-2 minutes, particularly evident for the low-energy channels shown in dark blue to green in Figure 5 (left, (a)). Such fluctuations are observed to be in good correlation with the magnetic field structuring (Figure 5 (a), (b), (c)), thus suggesting that particle acceleration is indeed taking place in an irregular fashion for this IP shock, where additional acceleration may be provided by the local magnetic structures. It has been previously shown that particle acceleration or re-acceleration can also occur via stochastic magnetic reconnection in dynamical small-scale magnetic islands confined inside magnetic cavities observed at 1 AU (Zank et al. 2015a; Zank et al. 2015b; Khabarova et al. 2015; Khabarova et al. 2016; Khabarova et al. 2017) and also at further distances in the heliosphere (Malandraki et al. 2019).

An interesting feature of the SolO observations is the two steep magnetic field enhancements observed in the 10 minutes upstream of the shock (Figure 5, right bottom panel) compatible with upstream shocklets typically arising from the interplay between shock-reflected, energetic ions, and upstream waves. Previous observations have shown that they are rarely observed at IP shocks, with only three such cases previously reported thus making this observation by SolO particularly interesting. Figure 5 right bottom panel, shows how the shocklets influence the low-energy electron population, using measurements by SolO-EAS. Shocklets are associated with rises in the electron density (magenta arrows in Figure 5, right bottom panel). The analysis of the pitch-angle distributions (PADs) of >70 eV electrons are found to be more isotropic in pitch-angle space in the vicinity of the shocklets, probably due to pitch-angle scattering induced by the compressive structures, in a similar manner as observed in the Earth's bow shock environment (Wilson et al. 2013). As Trotta et al. (2024) highlight this preconditioning of the upstream particle population, may have important consequences for electron injection and acceleration to higher energies (e.g. Dresing et al. 2022; Wijsen et al. 2023).

In a complementary multi-spacecraft study of the 2022 September 5 SEP event, Kouloumvakos et al. (2025) presented an observational and modeling analysis of the SEPs observed at PSP and SolO. Figure 5, right top panel shows the energy spectrogram of proton fluxes from the HET-Sun telescope on SolO at an energy range from ~10 MeV to ~90 MeV. The SEP event, at SolO, shows a prompt onset and a clear VD in the proton arrival times. From the VDA of the SEP measurements Kouloumvakos et al. (2025) highlighted that the SEP release was prompt at SolO but significantly delayed at PSP relative to the eruption onset. From the 3D reconstruction of the shock, this study shows that the shock apex propagated fast in the corona (<20 Rs) with a maximum speed at the apex of ~2480 km/s. Performing a 3D modeling of the shock and a numerical SEP transport simulation with the PARADISE model (Wijsen, 2020), this study demonstrated that PSP was initially magnetically connected to a weak shock region at the flanks of the CME, whereas SolO was connected to a stronger region near the nose of the CME-driven shock. The evolution of the shock characteristics at PSP --from subcritical to supercritical-- proved critical for the delayed SEP release and the observed inverse velocity arrival of SEPs above ~1 MeV. In contrast, SolO's connection to a strong shock region from the beginning of the event can explain the prompt SEP onset and intense SEP event recorded at that location. Moreover, the results of the SEP transport model seem to suggest that cross-field diffusion may not have played an important role in the SEP transport to PSP. Contrasting the results obtained from the SEP modeling for PSP and SolO, this study also showed a complex interplay between shock dynamics and particle acceleration. The authors concluded that, in spite of the fact that the PARADISE model effectively reproduced PSP observations, it overestimated SEP intensities at SolO, highlighting the need for more nuanced treatment of the efficiency of particle injection (Vainio et al. 2014) in future models.

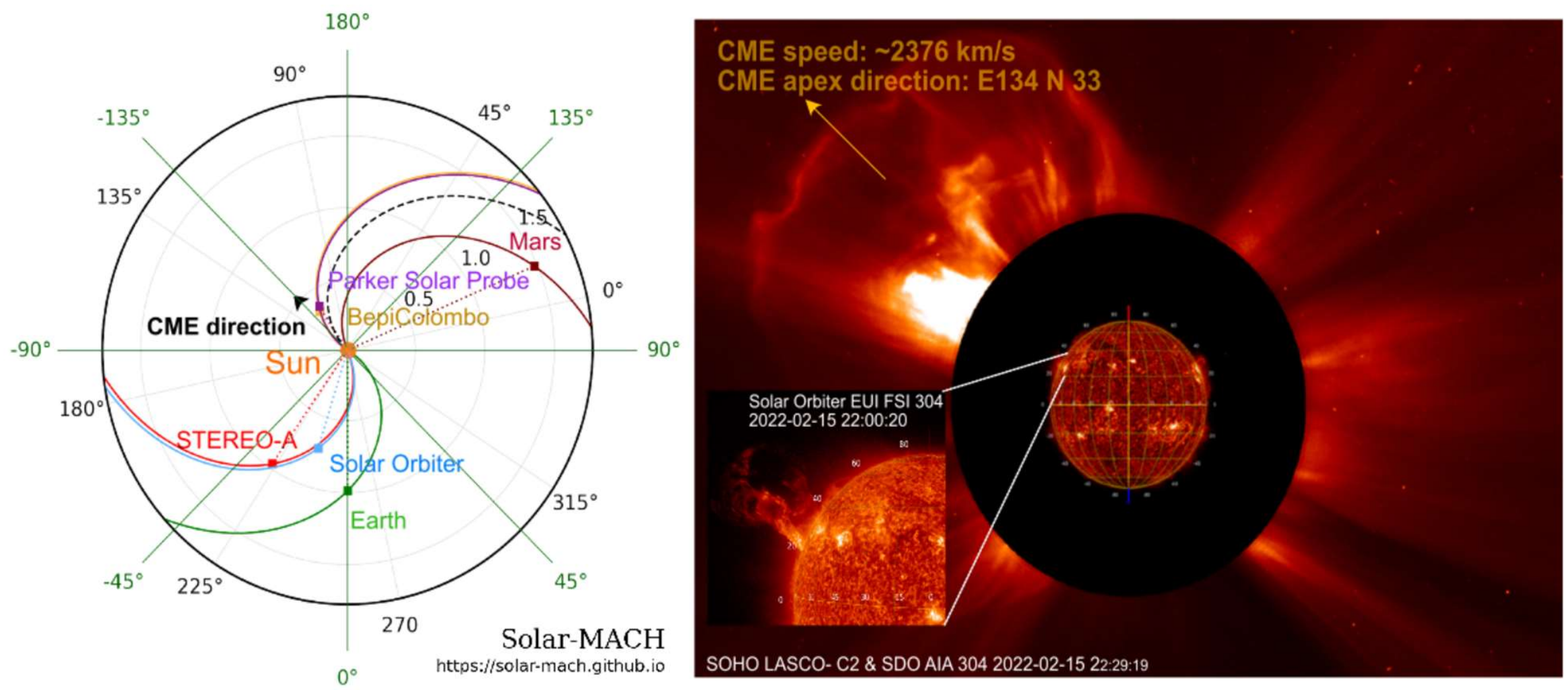

**Figure 6.** Overview of the spacecraft and planet configuration as well as solar activity on 2022 February 15. Left panel: Spacecraft and planet configuration in the inner solar system during the 2022 February 15 event, including BepiColombo, PSP, Mars, Earth, STEREO-A, and Solar Orbiter. The orbit plot is created in Stonyhurst coordinates with the Solar-MACH tool. Right panel: Overview of the solar activity on 2022 February 15, composed of images from SDO/AIA 304 Å and SOHO LASCO-C2 around 22:30 UT using JHelioviewer (Müller et al. 2017). It provides an overview of the corresponding CME from the erupting filament around 22:00 UT on 2022 February 15. The bottom left insert provides a better view of the erupting filament from the perspective of Solar Orbiter EUI FSI 304 Å. Based on the graduated cylinder shell (GCS; Thernisien et al. 2006, 2009) 3D model, the CME apex was heading in the direction E134 N33 with a speed of ~2376 km $s^{-1}$. (Reproduced from Khoo et al. 2024)

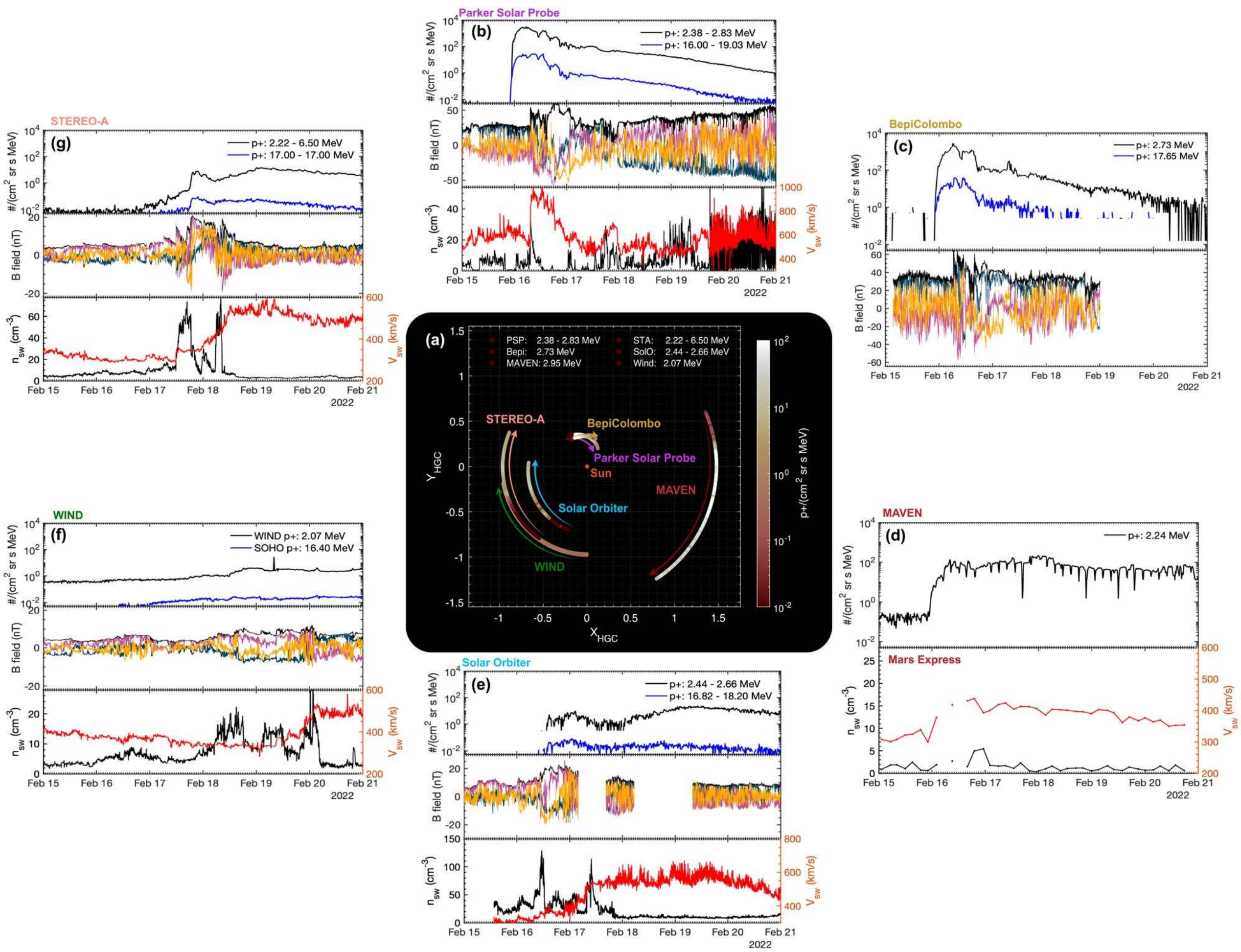

**Figure 7.** Overview of particle, magnetic field, and solar wind dynamics from a multi-spacecraft perspective. (a) Location of spacecraft in heliographic Carrington coordinates, where the color bar indicates the intensity variation of ~2 MeV protons observed by (b) PSP, (c) BepiColombo, (d) MAVEN/Mars Express, (e) Solar Orbiter, (f) Wind, and (g) STEREO-A (in the clockwise order) between 2022 February 15 and 21. We note that the WIND data are plotted with a slightly larger marker size to distinguish them from the STEREO-A data. The subplots show ~2 and ~17 MeV (when available) proton intensities from different spacecraft as well as measurements of magnetic field, solar wind velocity, and solar wind density when available. We note that proton fluxes from all spacecraft are in a time resolution of 15 minutes, and for (a), only points at every hour (or every fourth point) are plotted in order to reduce overlapping points. (Reproduced from Khoo et al. 2024)

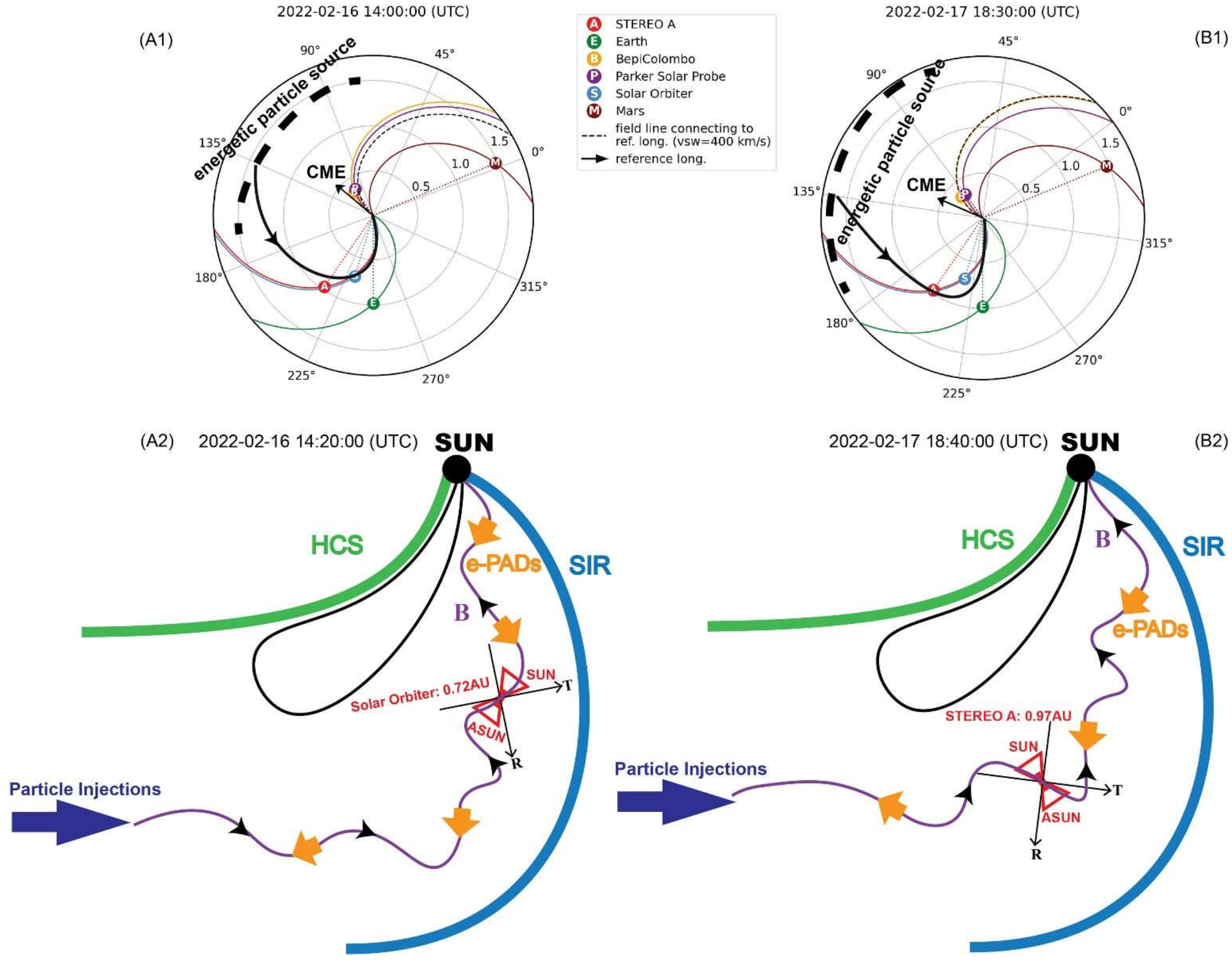

**Figure 8.** A sketch to describe Solar Orbiter (A1–A2) and STEREO A (B1–B2) observations; details are given in the main text. Figures (A1) and (B1) show the positions of Solar Orbiter and STEREO A, as shown in Figure 6. The CME propagation direction is marked by the black arrow, and the possible energetic particle source associated with the CME-driven shock is indicated by the bold dashed line. Figures (A2) and (B2) indicate the solar wind structures and particle features when Solar Orbiter and STEREO A encounter the focused anisotropy event at different locations, respectively. In the figures, an open small flux rope (SFR) (purple line) and a following closed loop (black line) are embedded between the SIR (blue line) and the HCS (green line). Moreover, the antisunward-flowing suprathermal electrons (gold arrows), magnetic field directions (black arrows), energetic particle injections (blue arrow), and the instrument viewing directions (red sketch) are displayed in the open SFR. (Reproduced from Wei et al. 2024).

## 3. Multi-spacecraft observations of a widespread SEP event on 2022 February 15-16

The SEP event of 2022 February 15–16, represents one of the most intense and spatially extensive SEP events observed in Solar Cycle 25, as analyzed by Khoo et al. (2024). It was associated with a fast CME that reached speeds of up to ~2376 km/s, as reconstructed using the Graduated Cylindrical Shell (GCS) model. This high-speed CME originates from an eruption that occurred behind the east limb of the Sun, likely associated with the old sunspot region 2936-2938 from the previous Carrington rotation. It drove a shock that propagated across a broad longitudinal range with an estimated shock speed of ~2315 km

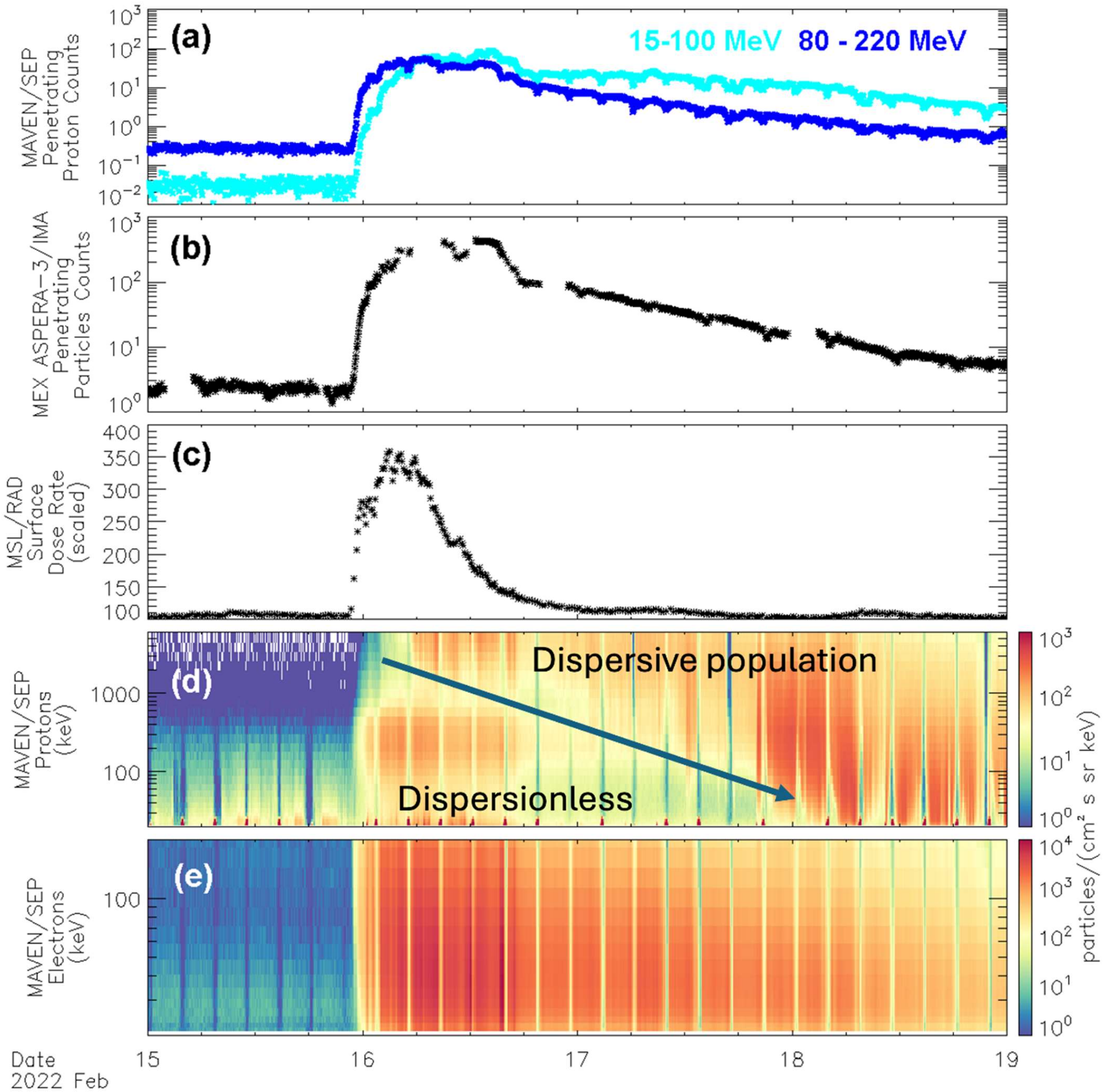

**Figure 9.** Particle dynamics on Mars. (a) Penetrating proton count rates from the SEP detector on board the MAVEN orbiter for 15–100 MeV protons (light blue) and 80–220 MeV protons (dark blue). (b) Penetrating particle count rates from the Mars Express ASPERA-3 IMA instrument. (c) Relative increase of Martian surface radiation dose rate scaled to the pre-event levels. Differential energy fluxes (colors) from MAVEN/SEP for (d) 20 keV to 6 MeV protons and (e) 20 to 200 keV electrons. The periodic drop in counts or fluxes seen in the MAVEN/SEP observations ((a), (d), and (e)) occur when MAVEN enters periapsis during its elliptical orbit around Mars. (Reproduced from Khoo et al. 2024)

$s^{-1}$ (±185 km $s^{-1}$, which is ±8% of the determined value; Kwon et al. 2014) at a heliocentric height of 25 $R_s$. Khoo et al. (2025) determined that the longitude and latitude of the CME apex were at -134° and 33° in Stonyhurst coordinates with a fixed tilt angle of -50° and a total angular width of ~58° in the ecliptic plane.

Its fortuitous timing and spacecraft configuration enabled a rare opportunity to study SEP propagation and acceleration across a wide longitudinal and radial range (0.35 au to 1.5 au; the closest being BepiColombo at 0.35 au) in the inner heliosphere (Figure 6). Observations were made by multiple spacecraft (Figure 7), including Parker Solar Probe, BepiColombo, Solar Orbiter, STEREO-A, near-Earth assets (SOHO, ACE, Wind), and Mars-based missions (MAVEN, Mars Express (Ramstad et al. 2018; Futaana et al. 2022)), and the Curiosity rover (Hassler et al. 2012)). However, due to the CME trajectory and tilt angle, these near-equatorial spacecraft likely encountered the CME flank, instead of the nose.

As highlighted by Khoo et al. (2025), there are several key features of this event. The 2022 February 15–16 SEP event exhibited a wide longitudinal enhancement, with enhancements across a wide range of magnetic footpoint separations, reaching up to ~165°, underscoring its classification as a widespread event. Spacecraft with footpoints magnetically well-connected to the eruption site—namely BepiColombo, PSP, and Mars—recorded substantially higher energetic proton fluxes than those observed by SolO, STEREO-A, and near-Earth missions such as Wind.

Notably, due to the proximity of the two spacecraft, BepiColombo and PSP observed a relatively similar particle-time intensity profile; both spacecraft observed energetic protons and electrons shortly after the Type III radio burst observation (that is succeeded by a Type II burst) at PSP and measured peak proton enhancements at shock arrival, suggesting a locally accelerated proton population by the shock. In contrast, SolO, STEREO-A, and near-Earth missions measured energetic protons approximately 1-day after the peak enhancement observed at PSP and BepiColombo. Solar wind measurements and WSA–ENLIL modeling revealed the presence of stream interaction regions (SIRs) at SolO, STEREO-A, and Earth during the SEP event, complicating the interpretation of particle sources (Figure 7). These overlapping structures challenge the separation of CME-driven and SIR-driven contributions to the observed SEP profiles.

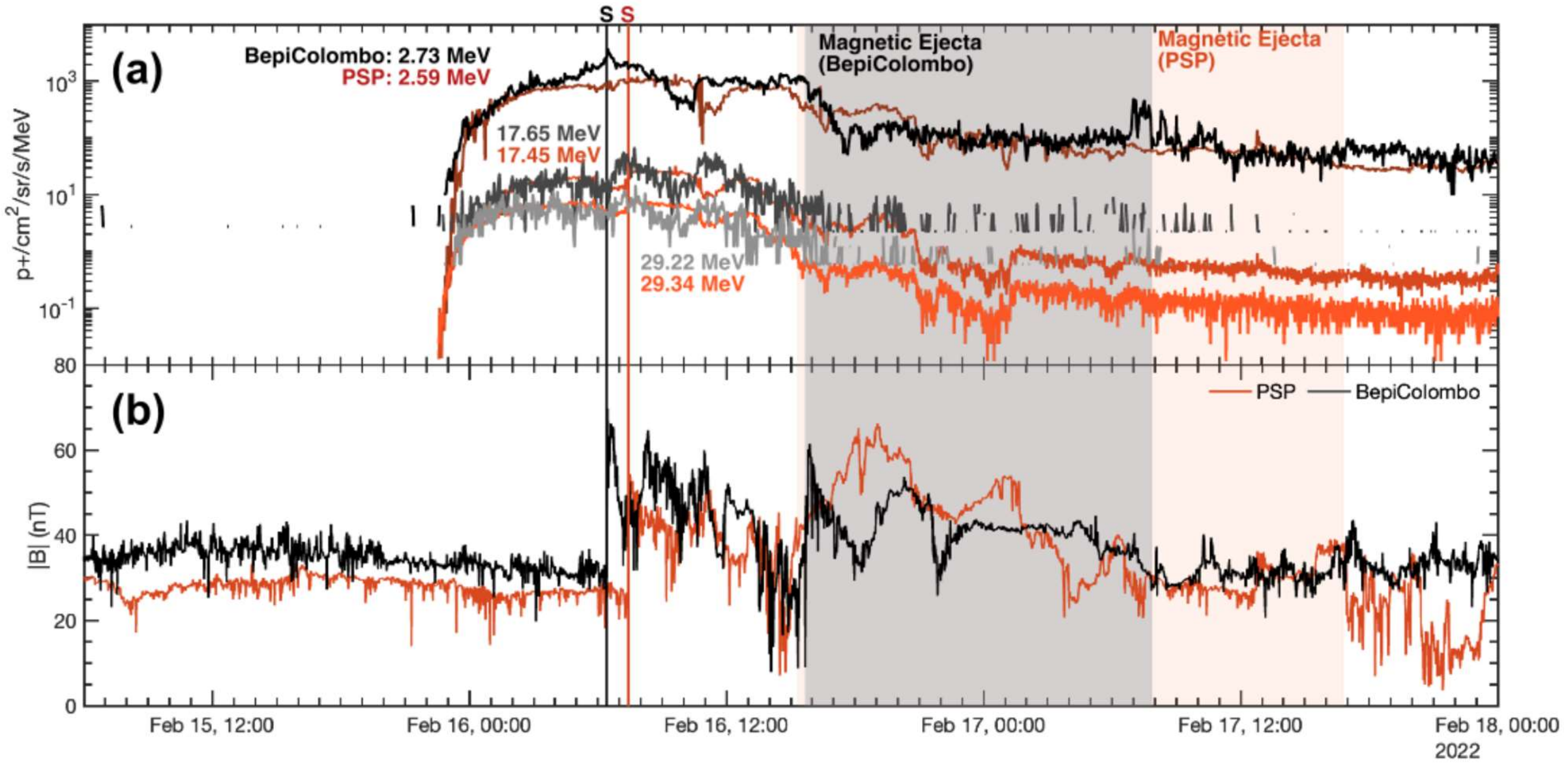

**Figure 10.** Comparison between PSP and BepiColombo measurements. (a) Comparison of one-minute resolution 2.59, 17.45, and 29.34 MeV proton fluxes from PSP (black/gray) with three-minute binned averaged 2.73, 17.65, and 29.22 MeV proton fluxes from BepiColombo (red/orange). (b) Comparison of the magnetic field magnitudes and their RTN components, respectively. Black and red solid lines represent PSP and BepiColombo data, respectively. The shaded gray and red regions indicate the magnetic ejecta observed on BepiColombo and PSP, respectively, while the black and red solid lines represent the shock arrival at BepiColombo and PSP. (Reproduced from Khoo et al. 2024).

Wei et al. (2024) examined the same SEP event on Solar Orbiter and STEREO-A in detail and reports very large and unexpectedly long-lasting anisotropies (on time scales of several hours) in energetic ion measurements observed by both spacecraft, which are nearly aligned along the same nominal Parker spiral (Figure 8). This energetic ion enhancement occurred during the rising phase of this widespread SEP event, but with a much delayed enhancement observed by STEREO-A. Their analysis suggests that this anisotropy has likely resulted from continuous injections of energetic ions from a well-connected particle source located beyond the orbit of STEREO A (Figure 8). It is noteworthy that Solar Orbiter also observed an interval of extremely large anisotropy dominated exclusively by sunward-streaming ions, which is interpreted as capturing the very early phase of ion injections onto newly connected magnetic field lines between Solar Orbiter and the particle source, likely making it the first reported event of this kind, based on our knowledge. This event also underscores the importance of studying anisotropies to uncover the sources, injection mechanisms, and transport processes of energetic particles.

The footpoints of BepiColombo, PSP and Mars were near the eruption site during this event (Khoo et al. 2024). MAVEN observed a similar time-intensity profile as the two near-sun spacecraft, signaling a good magnetic connectivity among these missions despite their “physical” angular distance. A closer look at the MAVEN measurement (Figure 9; Khoo et al. 2024) reveals two distinct proton populations during this period: 1) a dispersive population of 20 keV to 6 MeV protons (with strong VD observed by >1 MeV protons) that lasted for a few days and is likely associated with this CME; 2) a dispersionless population of 20 keV to ~1MeV protons that lasted for less than one day and is likely a result of a merged CME that erupted prior to this event.

Additionally, the Mars Ground Level Event (MGLE) observed is another key feature of this event. During this event, the Curiosity rover on Mars’ surface also detected the first ground-level enhancement in Solar Cycle 25 and one of the largest MGLE events recorded since its operation (Posner et al. 2025). Such an MGLE event is likely associated with >150 MeV protons, as discussed in Guo et al. (2019), and Zhang et al. (2024). This SEP event was also observed at Mars by Tianwen-1 orbiter/Mars Energetic Particle Analyzer (MEPA) instrument (Tang et al. 2020). This enables studies like Zhang et al. (2024) to confirm the instrument capability to measure energies up to 100 MeV and demonstrate its complementary role to MAVEN in advancing our understanding of Mars’ environment during a major solar proton event.

Furthermore, the proximity of BepiColombo and PSP during this event provided an excellent opportunity for cross calibration between the BepiColombo/BERM and PSP/EPI-Hi instruments. Khoo et al. (2024) carried out a detailed comparison between the BepiColombo/BERM and PSP/HET, finding that the proton fluxes between 2-30 MeV are generally consistent. However, BepiColombo/BERM intensities were ~35% lower than those from PSP/HET, yielding an intercalibration factor of ~1.35 (see Figure 10 and Appendix D in Khoo et al. 2024). Since PSP/HET and PSP/LET are generally well-calibrated with each other (Joyce et al. 2021), this result also suggests that BERM is likely within a reasonable agreement with LET.

This alignment also enabled a detailed examination of the CME’s mesoscale structure using BepiColombo and PSP data, as discussed in Palmerio et al. (2024). They conducted a very thorough analysis, comparing key CME-related features—including shock parameters, sheath regions, and CME ejecta—and revealed both similarities and striking differences (Figure 11).

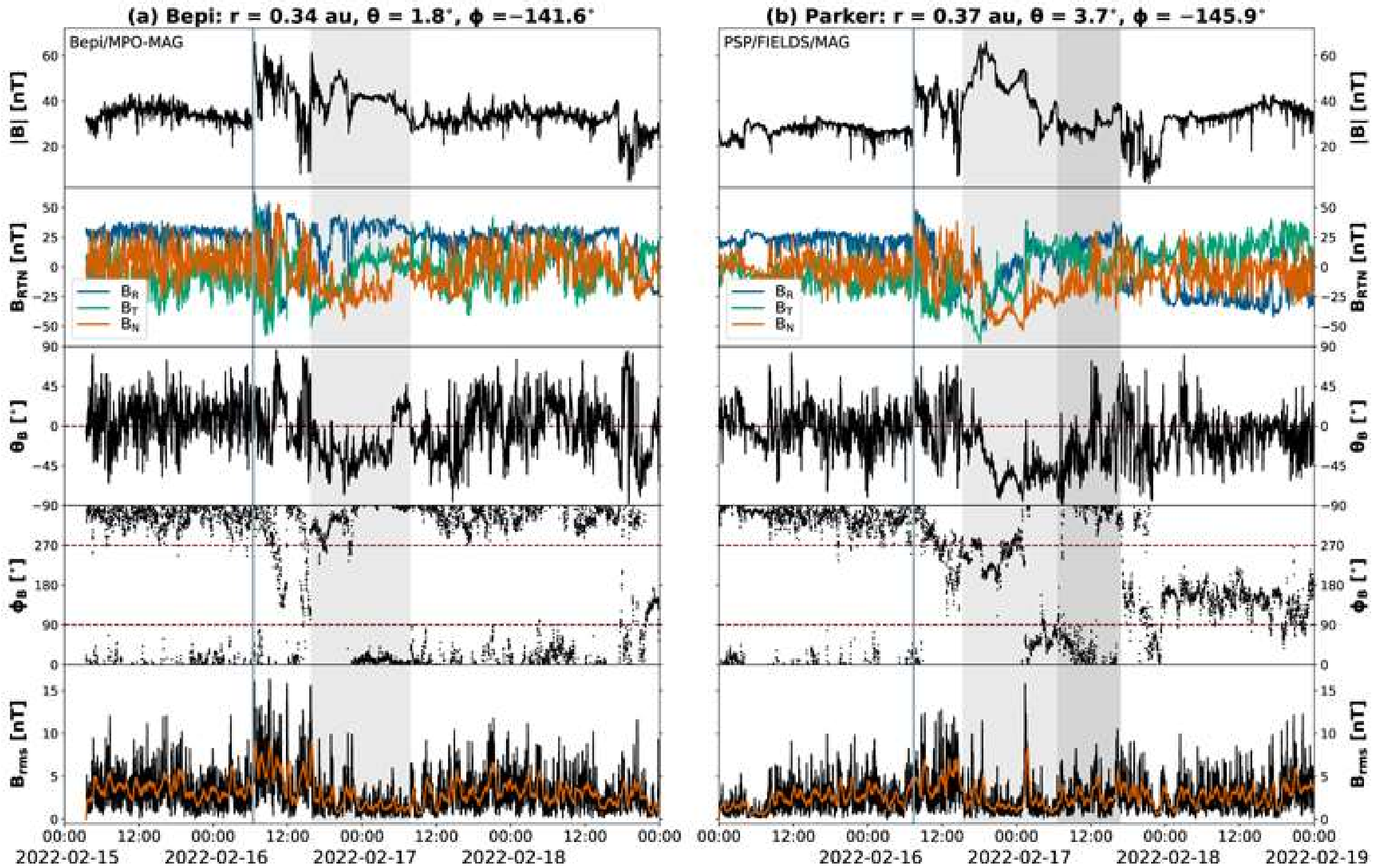

**Figure 11.** Magnetic field measurements (1 minute cadence) at (a) BepiColombo and (b) PSP for the period 2022 February 15–18. Each plot shows, from top to bottom: magnetic field magnitude, magnetic field Cartesian components in RTN coordinates, latitudinal angle of the magnetic field, longitudinal angle of the magnetic field, and rms error of the magnetic field (using a rolling average over 15 minutes for each component, the superposed orange curve displays the 30 minute rolling average of the same data). The (r, θ, ϕ) values displayed at the top of the panels for both spacecraft refer to 2022 Feburary 16, 07:00 UT, and the angles are reported in Stonyhurst coordinates. The vertical gray lines indicate the arrival of the CME-driven shock at either spacecraft, while the gray shaded areas highlight the CME ejecta. At Parker, the full ejecta (magnetic obstacle) includes both gray regions, and the flux rope (magnetic cloud) structure within it is represented by the lighter gray area. (Reproduced from Palmerio et al. 2024).

For example, both spacecraft observed an IP shock with comparable magnetic compression, yet PSP detected a moderately quasi-parallel shock, whereas BepiColombo yielded two conflicting and likely unreliable results (one strongly parallel and one strongly perpendicular). In the sheath region, measurements from both spacecraft indicated similar magnetic environments, such as comparable magnetic field fluctuations and the unusual presence of a planar magnetic structure (PMS) in the middle of the sheath. However, significant differences emerged in various areas. For instance, BepiColombo observed a relatively coherent CME ejecta lasting ~16.2 hours, while PSP recorded a ‘two-part’ ejecta extending ~25.5 hours. These pronounced local variations within the large-scale structure underscore the existing knowledge gap regarding CME mesoscale variability.

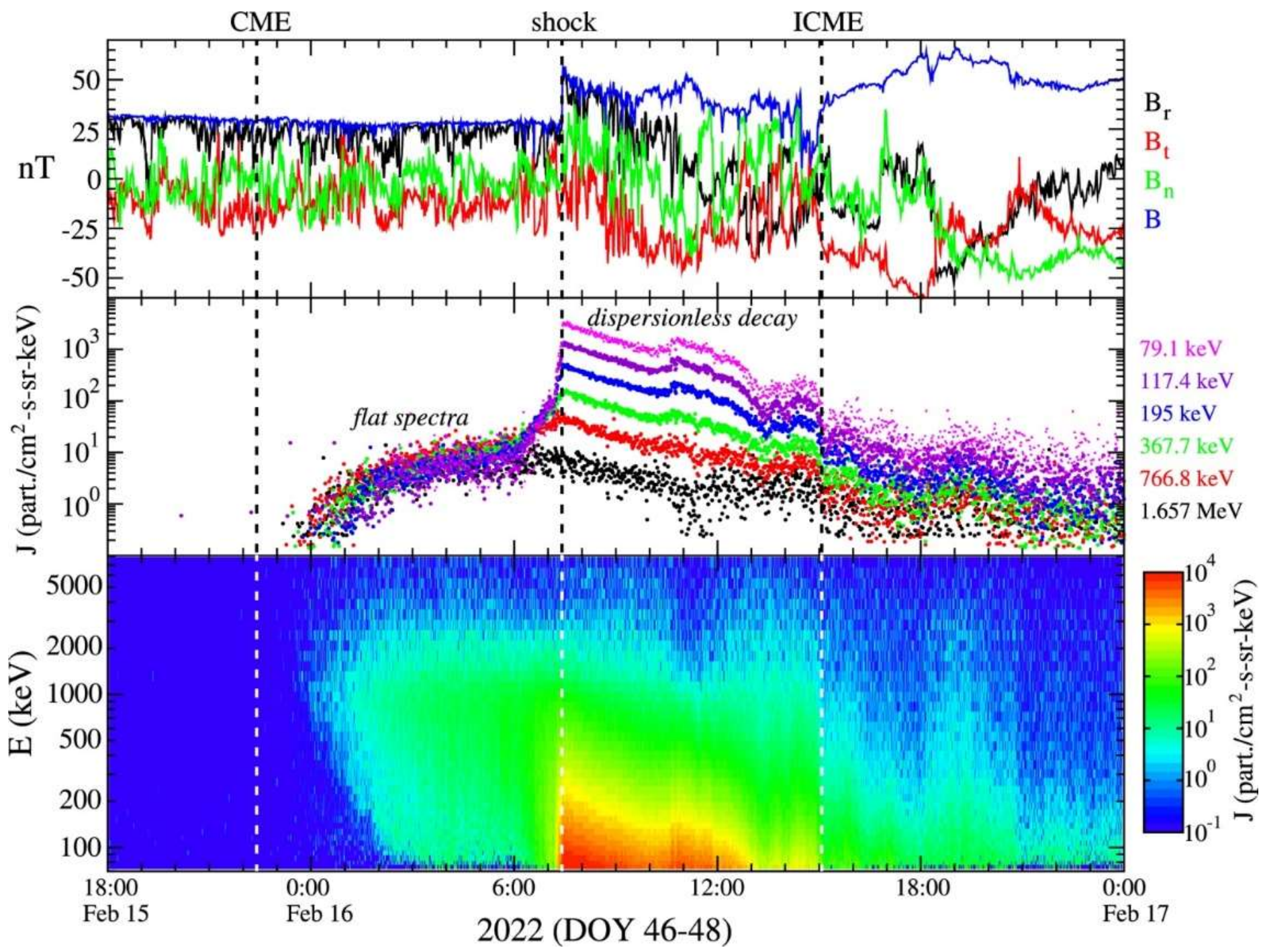

**Figure 12.** Overview of the ESP event observed at PSP. From top to bottom: 1-s resolution magnetic field vector from the PSP/FIELDS instrument, 1-minute resolution flux of protons at 6 different energies from the PSP/ IS⊙IS instrument, and the flux of protons over a broader range of energies in a color spectrogram format. The first vertical dashed line at the left represents the time in which STEREO-A/Cor2 first observed the CME, the second dashed line, in the middle, is at the time of shock passage. The last dashed line, at the right, represents the approximate time where the magnetic flux rope associated with the Interplanetary CME (ICME) first passes PSP. (Reproduced from Giacalone et al. 2023).

### 3.1 *In-situ analysis of the ESP event on 2022 Feb 16 seen by PSP*

During this widespread event, PSP also observed an energetic storm particle (ESP) event during the shock passage at PSP. Giacalone et al. (2023) performed several analyses of this event. In particular, they focused on the flux and spectra of ~0.5-5 MeV/nuc energetic ions, using data from the PSP/IS⊙IS /EPI-Lo instrument. As shown in Figure 12, the event was characterized by a large increase in the fluxes of these ions several hours before the arrival of the shock, and a few hours after the CME was observed in the ST-A Cor2 images, followed by a gradual increase in particle fluxes until about 15-30 minutes prior to the shock arrival during which time the fluxes increased dramatically, and peaked at the time of the shock crossing. Immediately after the shock, the proton fluxes decreased nearly exponentially. During the slow rise of particle fluxes, in advance of the shock, the intensity was nearly the same at all energies - this leads to a nearly flat energy spectrum. During the rapid rise phase, the increase had an energy dependence, and was nearly power law at the shock itself. The exponential decay rate was the same at all energies behind the shock. Giacalone et al. (2023) used this data to determine spatial diffusion coefficients as a function of energy far from the shock, and near the shock. They found that the diffusion coefficient was nearly independent of energy far from the shock, and had a nearly linear dependence on energy near the shock. The diffusion coefficient far from the shock was also much greater than that near the shock. They also made a determination of the source of the energetic particles by analyzing the increase in the intensity above the background, seen far before onset of the event, and comparing it with predictions of diffusive shock acceleration theory. They concluded that the source of energetic particles was the solar wind and not the re-acceleration of suprathermal particles. They also focused on the unusual exponential decay of particles behind the shock, concluding that the CME was in a stage of over-expansion during the time it crossed PSP, and resembled a blast wave, instead of a driven shock. In this case, the exponential decay was caused by the particles filling a rapidly increasing volume.

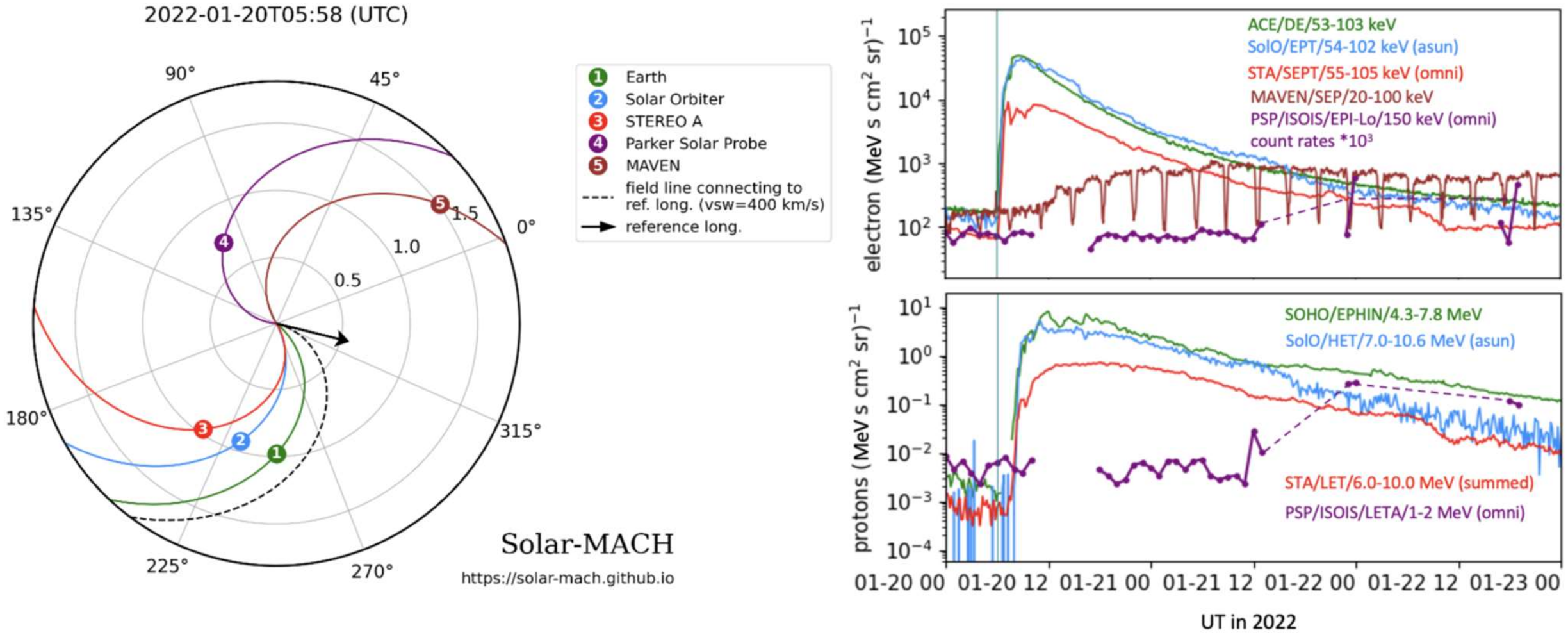

**Figure 13.** Left: Longitudinal spacecraft configuration and magnetic connectivity at 05:58 UT on 2022 January 20 using the Solar-MACH tool in the Carrington coordinate system. Right: Multi-spacecraft SEP measurements. The blue vertical line indicates the time of the soft X-ray peak of the flare (~5:58 UT) associated with the SEP event. (Reproduced from Rodríguez-García et al. 2025)

## 4. 2022 January 20 widespread SEP event: SEPs injected inside and outside a magnetic cloud

Rodríguez-García et al. (2025) carried out an extensive study of the widespread SEP event observed on 2022 January 20, in order to identify its solar source and to investigate the acceleration and propagation conditions that could affect the observed SEP properties at the different but closely spaced observers, in particular at the location of SolO. The SEP event was found to be associated with an M5.5 flare erupting from AR 12929 located at Stonyhurst N08W76 peaking at 05:58 UT on 2022 January 20, as well as a fast CME-driven shock of ~1433 km $s^{-1}$. Figure 13, left shows the spacecraft locations in the heliographic equatorial plane along with nominal Parker field lines at 05:58 UT on 2022 January 20, connecting each spacecraft with the Sun in the center of the plot, using measured solar wind speeds when available. The black arrow marks the longitude of the associated flare and the dashed spiral depicts the nominal magnetic field line connecting to this location. The upper panel of Figure 13 right shows near-relativistic (~20-150 keV) electron intensities whereas the lower panel shows ~5 MeV proton intensities as measured by the different spacecraft, with the same color-code as shown in the left. SolO, STEREO-A, near-Earth spacecraft as well as MAVEN (only electrons) observed the SEP event. PSP shows a later increase which may be related to a later eruption from a different AR. Thus, based on the available observations Rodríguez-García et al. (2025) argued that the SEPs accelerated by this solar event resulted in a particle spread over at least 160° in the heliosphere, from STEREO-A to MAVEN. For each spacecraft Rodríguez-García et al. (2025) inferred the magnetic connectivity between the spacecraft and the source surface by means of the nominal Parker spiral solutions using their heliospheric distances and the observed solar wind speeds. They further tracked magnetic field lines downward to the photosphere using the Potential Field Source Surface (PFSS) coronal field model and derived the instantaneous connectivity, concluding that the connectivity of the near-Earth spacecraft, SolO and STEREO-A to the solar surface is very similar between the three spacecraft, i.e. ~316° longitude and ~-16° latitude.

Then, the difference with the solar flare region is of ~9° in longitude and ~24° in latitude for the three aforementioned spacecraft. Rodríguez-García et al. (2025) focused their study on these near-1 AU observations of the SEP event.

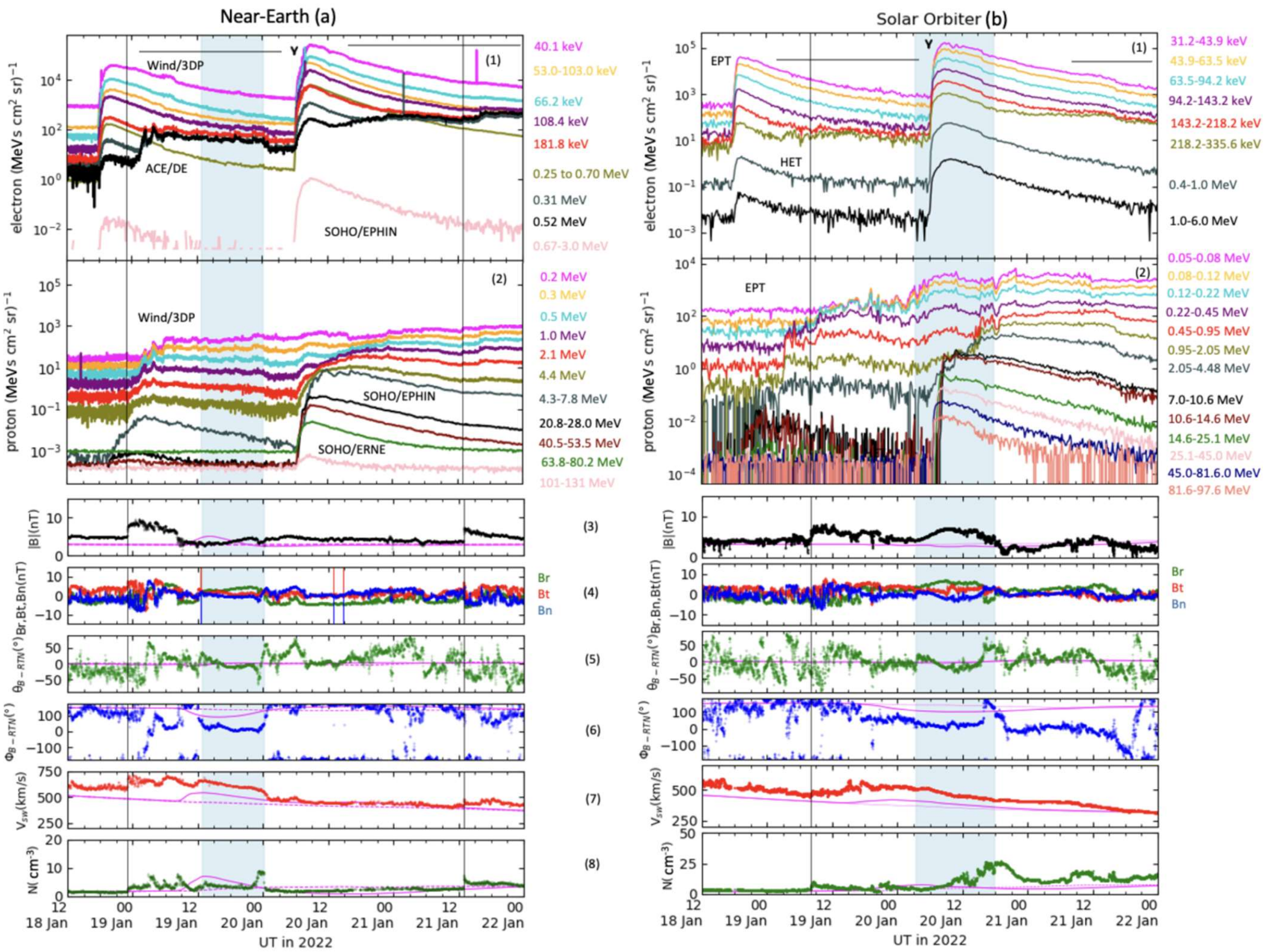

**Figure 14.** In situ SEP time profiles, plasma and magnetic field observations (in RTN coordinates) by near-Earth spacecraft (a) and SolO (b) (see text). Solid vertical lines indicate IP shocks, whereas the blue shaded areas indicate the MC. Periods of proton contamination in the Wind/3DP (SolO/EPT) instrument measurements are indicated with a horizontal line in the left (right) panel (1). (Reproduced from Rodríguez-García et al. 2025)

Figure 14 shows in situ energetic particle intensity time profiles, plasma and magnetic field observations by near-Earth spacecraft (a) and SolO (b). The first two panels of Figure 14(a) show the SEP event on January 20 as measured by the Sun-directed telescopes of the various instruments. The arrow head at the top of panel 1 indicates the onset of the parent solar flare. A resulting fast rise of energetic electrons reaching energies of at least 0.7 MeV are observed, with the proton-intensity also showing a fast ion intensity increase for 4-100 MeV energies, with a gradual increase for the lower ion energies. Figure 15 presents energetic particle anisotropy information from the three spacecraft, utilized for the study of the energetic particle pitch-angle distributions (PADs). As shown in panel 1 of Figure 15 (electrons) and Figure 6 of Rodríguez-García et al. (2025) (protons, not shown here) electrons and protons arrived to the near-Earth environment travelling away from the Sun with a pitch-angle of 180°, as the magnetic field polarity is negative during this period (Br negative, as shown in panel 4 of Figure 14(a), in RTN coordinates, corresponding to a sunward directed field).

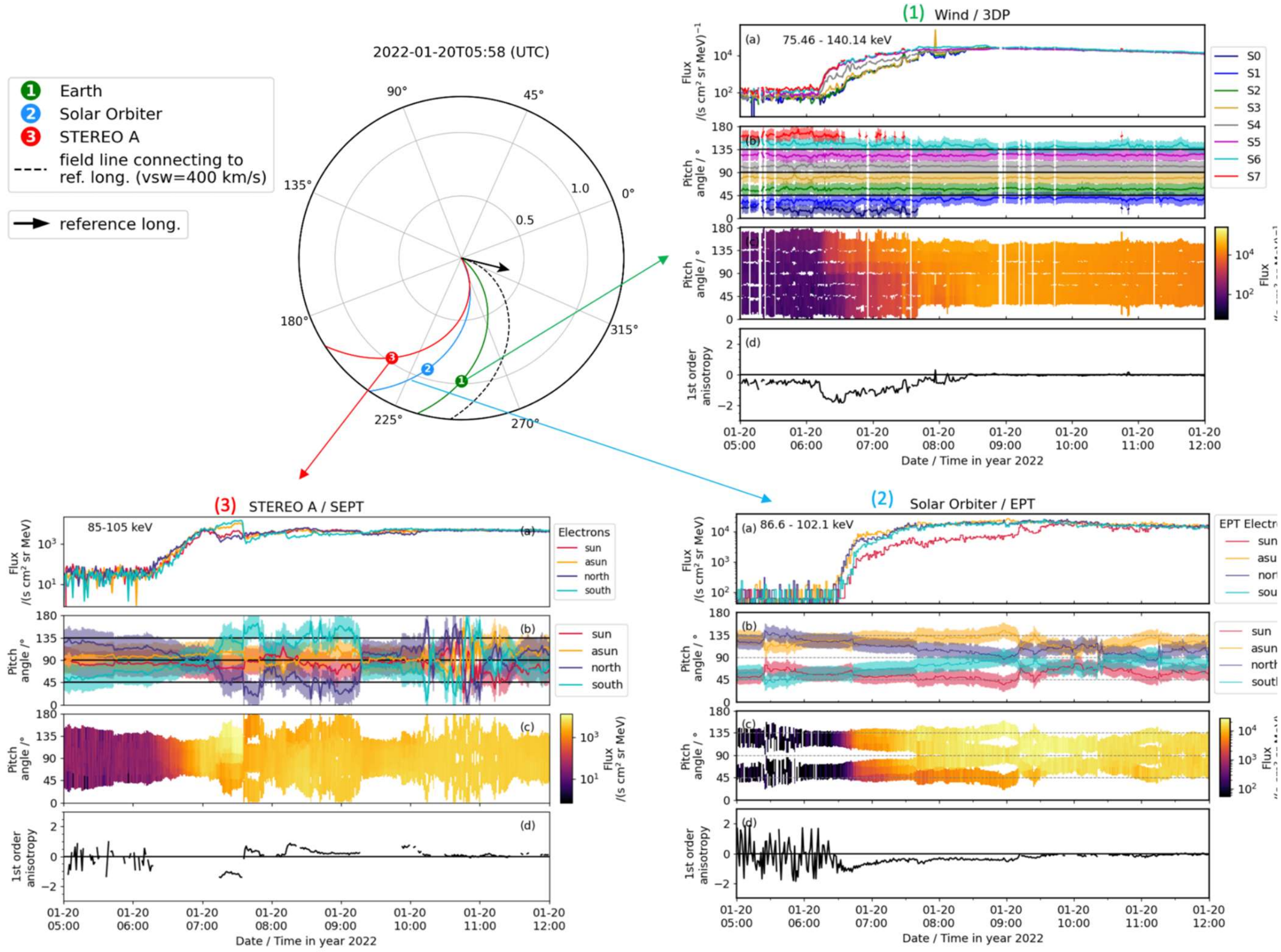

**Figure 15.** Electron pitch-angle distributions (PADs) measured by Wind (1), SolO (2) and STEREO-A (3). The top left panel shows the longitudinal spacecraft configuration and nominal magnetic connectivity at 05:58 UT on 2022 January 20. (Reproduced from Rodríguez-García et al. 2025).

In the near-Earth spacecraft, the arrival of an IP shock (first vertical line in Figure 14(a)) on January 18 is observed, followed by a sheath region with increased magnetic field and large fluctuations in the magnetic field and a subsequent region with coherent magnetic field rotation. Based on these observations, Rodríguez-García et al. (2025) identified an ICME indicated by the blue shaded area, starting at 12:40 UT on January 19 lasting until 00:11 UT on January 20, that had recently left the near-Earth environment. This ICME is found to be the same structure arriving later at SolO as shown in Figure 14(b), the top two panels of which show the SEP event observed by SolO. The data correspond to the particles measured by the anti-sunward looking telescopes, which measured the earliest onsets and highest intensities as shown in Figure 15 (2) and Figure 7 of Rodríguez-García et al. (2025). The electron event (panel 1) shows a fast intensity increase to energies above ~1 MeV in both EPT and HET measurements, whereas the energetic ion observations (panel 2) by SolO/EPD/HET show a clear velocity dispersion reaching energies up to ~80 MeV. As shown by the magnetic field and plasma data in panels 3-8 of Figure 14(b), using

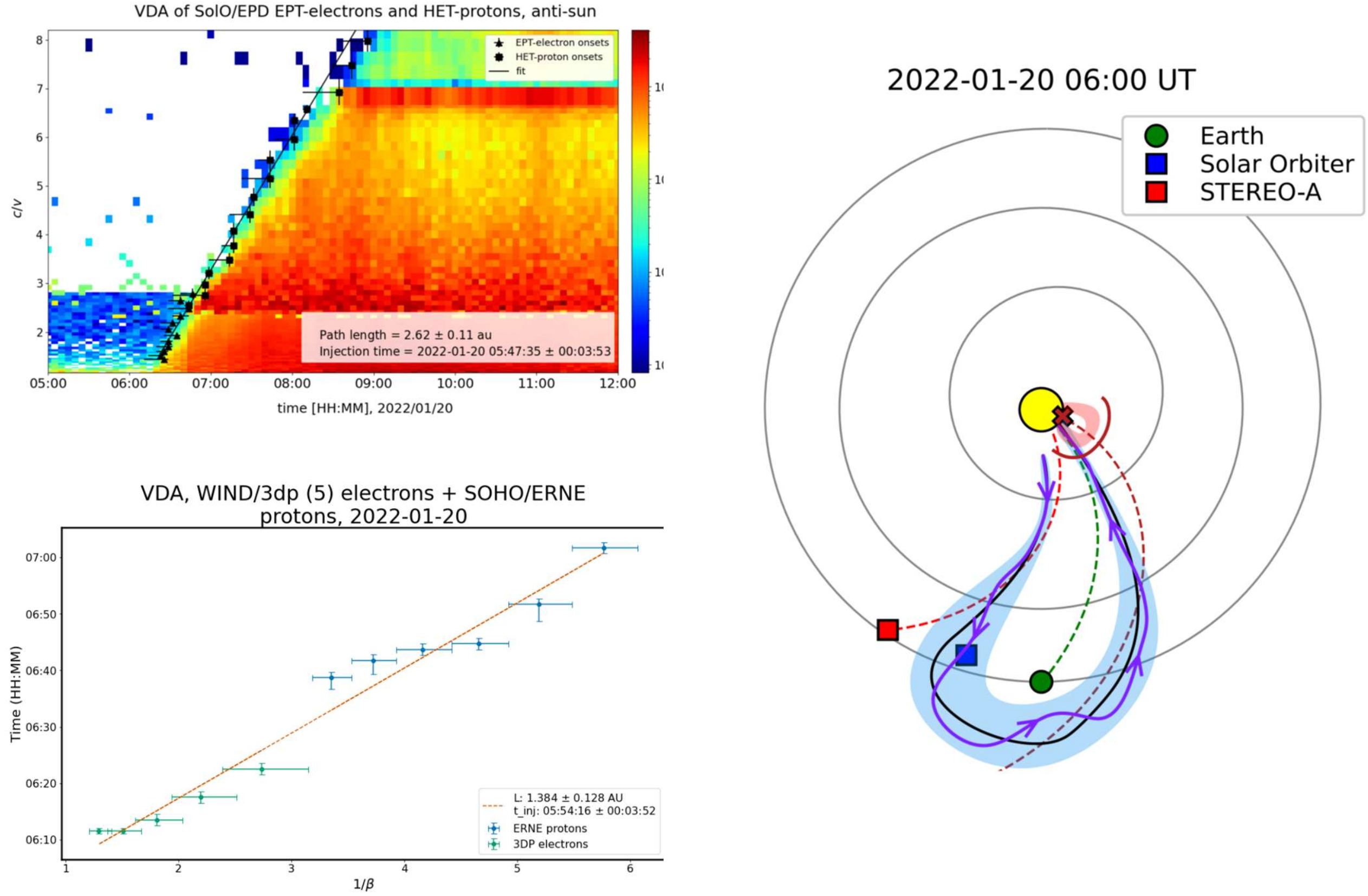

**Figure 16.** VDA results of the onset of the SEP event at SolO (top left panel) and near-Earth spacecraft (bottom left panel). Right: Sketch of the interplanetary configuration of the 2022 January 20 SEP event (see text). Reproduced from Rodríguez-García et al. (2025).

SolO/MAG and SolO/SWA data, the SEP event onset is observed to take place during the passage of an ICME over SolO. An IP shock is impacting the spacecraft at 08:02 UT on January 19, indicated by the vertical line in Figure 14(b). A Magnetic Cloud (MC) arrives at 03:28 UT on January 20, just before the particle event onset, observed until 17:52 UT (blue shaded area). The energetic electrons and the higher energetic ions (> 7 MeV) are propagating inside the ICME. This ICME has been found to be most likely associated with a CME that appeared in LASCO C2 field of view at 20:48 UT on 2022 January 16, associated with a flare erupting four days earlier from the same AR 12929 as the one related to the SEP event on January 20. Near-relativistic electron fluxes and anisotropy characteristics have been previously utilized as tracers of the large-scale structure and topology of the IMF embedded within ICMEs in and out of the ecliptic plane, providing unique clues about whether the detected ICME structure has been detached from the solar corona, or still magnetically anchored to it (e.g. Malandraki et al. 2001; 2002; 2005; Gómez-Herrero et al. 2017). A noteworthy feature is the negative anisotropy index shown in panel d of Figure 15 (2) for SolO, corresponding to particles propagating towards the Sun, as the local magnetic vector is pointing outward during this period (shown in panel 4 of Figure 14(b)). A similar PAD is observed for the proton intensities, indicating particles propagated towards the Sun. These sunward-directed fluxes for the first arriving particles at SolO is quite an unusual feature. It is noteworthy that this feature is in agreement with the PAD of the solar wind electrons (not shown) which show that at the time of the SEP onset only the anti-parallel beam is observed. This is in agreement with the solar wind electrons propagating towards the Sun inside the western (longer) leg of an ICME connected to the Sun and the eastern (shorter) leg being disconnected, as shown in the detailed sketch in Figure 16, right.

Figure 16, left, shows VDA results for SolO (upper panel) and near-Earth spacecraft (lower panel). An effective particle propagation length of L = 1.4 ± 0.1 AU for near-Earth observers, which is close to the nominal Parker spiral length for near-Earth (~1.08 AU) using the measured solar wind speed. However, for SolO the effective length travelled by the particles is calculated to be L = 2.6 ± 0.1 AU, i.e. much longer than the length of ~0.99 AU expected for a nominal Parker spiral field estimated with the measured solar wind speed and scatter-free propagation. This indicated a possible non-standard IMF topology, with the long effective path length to be in agreement with the particles propagating inside an ICME to arrive to the SolO location.

From this VDA the estimated injection time of the particles were derived and found to be similar within uncertainties for the three spacecraft, near-Earth, SolO, and STEREO-A. Comparison with radio and hard X-ray (HXR) signatures showed that the estimated injection times are co-temporal with the presence of nonthermal HXR peaks, type IIs, and type IIIs starting at 80 MHz. Furthermore, via 3D reconstruction of the CME-driven shock the first time the shock wave intersects the magnetic field lines connecting to the near-Earth, SolO and STEREO-A spacecraft was estimated and compared with the VDA results. The connection to the shock is found to be in agreement and co-temporal, with the particle injection time for the three spacecraft, considering the time uncertainties. In addition to the particle composition features observed typical of gradual events during this event, as well as the presence of type III bursts starting at 80 MHz being co-temporal with the observed type II burst, suggest that the CME-driven shock is likely the main accelerator of the particles.

Figure 16, right, shows the sketch of the interplanetary configuration of the 2022 January 20 SEP event as concluded by Rodríguez-García et al. (2025). Earth, Solar Orbiter, and STEREO-A are indicated by the green circle and the blue and red squares, respectively. The ICME associated with the CME that erupted on 2022 January 16 is shown in blue. The CME and the CME-driven shock corresponding to the SEP event on January 20 are represented by the red shading and the red curve, respectively. STEREO-A and Earth are connected to the solar source through the nominal Parker spirals indicated with the dashed coloured lines using the measured solar wind speeds. SolO is embedded in an ICME, shown with a blue shading, with the axial magnetic field in black. The longer leg of the ICME was still anchored to the solar surface, based on the solar wind electron PAD. The solar source identification indicates that this leg is connecting to AR12929, and connected to the EUV wave and CME-driven shock related to the particle event, indicated as a red curve. The determined electron propagation path length inside the MC is found to be around 30% longer than the estimated length of the loop leg of the MC itself which is consistent with low number of field line rotations calculated (N≤6).

Thus, pre-existing structures in the heliosphere and the IMF are deemed to be very important for the transport and spread of SEPs (Richardson and Cane, 1996). During the 2022 January 20 SEP event, energetic particles are injected over a wide angular region into and around a pre-existing magnetic cloud (MC) structure ejected on 2022 January 16, which was present in the heliosphere at the time of particle onset. The sunward-propagating particles measured by SolO are interpreted as resulting from particle injection into the western (longer) leg of the MC, which remains magnetically connected to the Sun, followed by transport along the large-scale MC structure toward the spacecraft located in the eastern leg.

This scenario is supported by several observational constraints. In particular, the particle onset timing is consistent with an early connection between the EUV wave and the western flank of the MC, suggesting that the initial acceleration and release of particles occurred in this region rather than at later times when the shock could have expanded toward the eastern leg. In addition, VDA yields a relatively long effective path length (L = 2.6 ± 0.1 AU), which is difficult to reconcile with a more direct injection scenario near the ICME nose. Instead, such an extended path length is naturally explained by particle propagation along the structured magnetic field of the MC, from the western leg toward the eastern leg where SolO was located.

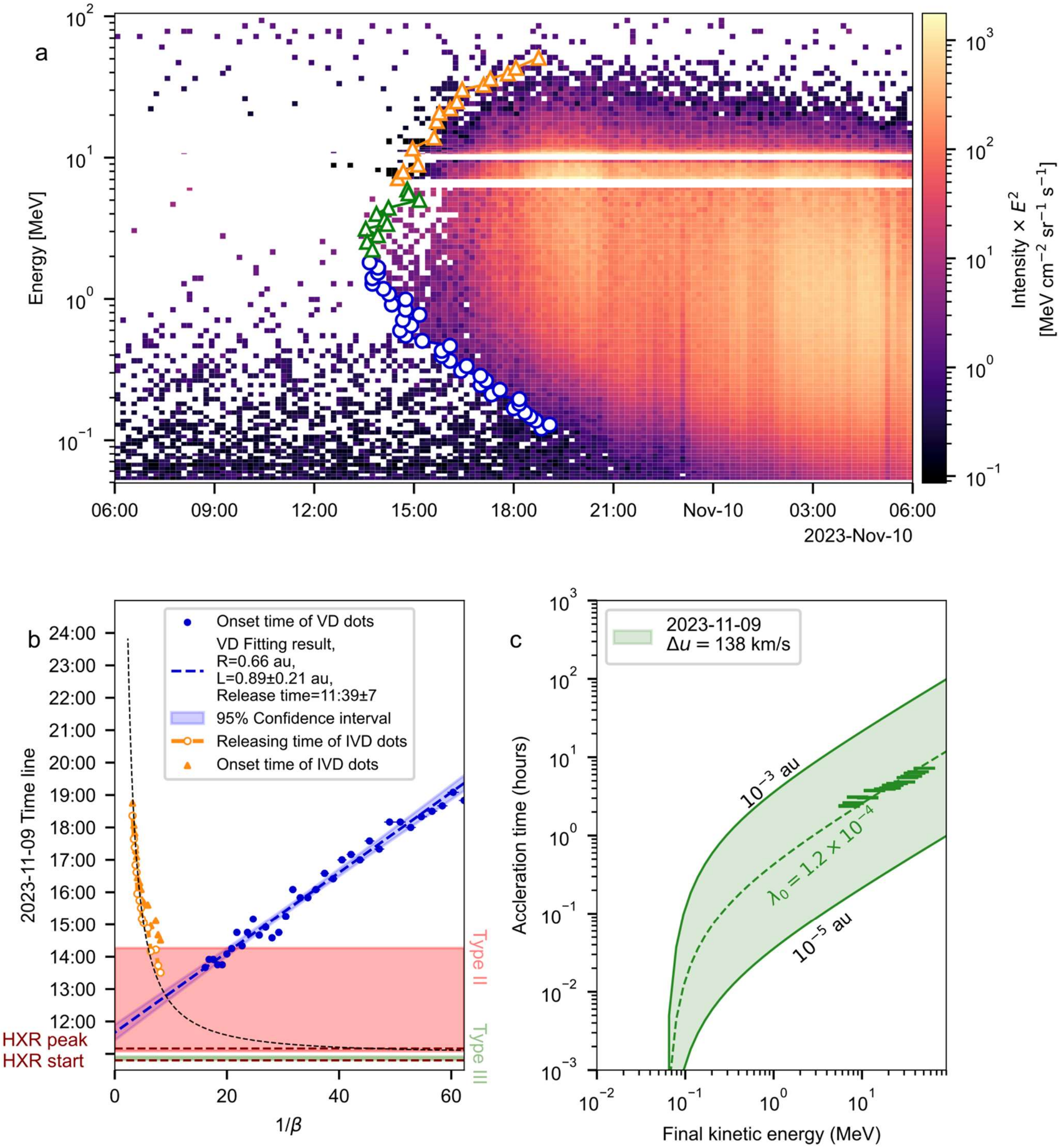

**Figure 17.** Panel a: dynamic spectrum of 2023-11-09 event. The onset times of particles exhibiting VD are marked with blue circles, while particles exhibiting IVD are marked with orange triangles. The transition energy is around 3 MeV, marked by green triangles. Panel b uses the onset times of VD particles in panel a (blue dots), to derive the release time and path length of protons below about 2 MeV (as shown in the legend). Orange triangles are the onset times of higher-energy IVD particles versus energy (represented by $1/\beta$ which is $c/v$ on the x-axis), and the orange circles are the derived release time which increase with energy. The DSA-fitted release time is plotted as the dotted line. Panel c: $\tau_a$ versus the final proton energy. The colored band corresponds to the possible solutions of $\tau_a$ with $\lambda_0$ ranging from $10^{-5}$ to $10^{-3}$ au. The markers denote the observed acceleration time derived from the observations based on the IVD analysis. Adapted from Li et al. (2025).

## 5. Investigation of the delayed arrival of faster SEPs observed by SolO: Observations and modeling

The long-standing picture of SEP events has been that faster particles arrive earlier, producing the well-known VD pattern in dynamic spectra. This behavior has commonly been interpreted as evidence that particles of different energies are released nearly simultaneously and propagate along comparable magnetic-field trajectories in a nearly scatter-free way. The observational examples discussed in the preceding sections, however, show that SEP timing and intensity profiles can be strongly affected by the evolving properties of CME-driven shocks, by the observer's magnetic connection to the acceleration region, and by the structured heliospheric environment through which the particles propagate. A particularly striking example is the 2022 September 5 event observed by PSP, discussed in Section 2.2, in which particles below about 1 MeV showed normal velocity dispersion whereas higher-energy particles arrived later, producing an "inverse velocity dispersion" (IVD) signature. In the same event, however, Solar Orbiter observed normal velocity dispersion at 0.7 au, indicating that the inverse-arrival signature depended on the observer's location and magnetic connection. This contrast motivates a more comprehensive examination of delayed high-energy SEP arrivals from the perspective of shock acceleration and particle transport. Recent Solar Orbiter observations of multiple IVD events provide such an opportunity, revealing IVD signatures that have been investigated through both observational analyses and modeling.

Li et al. (2025) selected and catalogued 10 events which they have termed as IVD events, observed by SolO from early 2022 to mid-2024 at heliocentric distances of 0.49–0.95 au; three with complete remote-sensing observations (2023-11-09, 2023-12-24, 2023-12-31) covered by Earth-view and STEREO-A spacecraft were investigated in depth. Taking the 2023-11-09 SEP event as an example, two C-class flares and a slow halo CME (~870 km/s) were associated with the event. In-situ particle data measured by SolO at 0.66 au show the event onset at 13:32 UT with the first-arrival energy at around 3 MeV, as shown in Figure 17(a). For the VD part (E<~2 MeV), a standard VDA was applied. VDA results are shown by blue dots and dashed line in Figure 17(b) and give the release time $t_0$ = 11:39 ± 7 min UT, broadly consistent with the flare hard-X-ray peak and type II start (~11:05 UT), indicating that these particles were likely accelerated by the flare process or the initial phase of the shock.

For the IVD part (E'>~7 MeV), a procedure is considered where particles of different energies were released at different time $t_{release}(E')$ (as the shock propagates outward) and they also arrived at the observer with different path lengths $L(E')$, as described in the equation below,

$$t_{onset}(E') = t_{release}(E') + \frac{L(E')}{v(E')} \qquad (1)$$

The above equation cannot be solved directly, but an iterative approach is used to derive the best-fit $t_{release}(E')$ and $L(E')$ accounting for outward motion of the CME shock (shortening the particle path) with a drag-based model (DBM, Dumbović et al., 2021). This yields energy-dependent release times that increase with energy, i.e., the higher-energy protons were released later, when the shock was already at ~0.05–0.2 au. This indicates that these particles were likely accelerated by the shock as it traveled outward. Using the first three hours after the onset in each energy band (to minimize transport effects that are amplified at later stages of SEP events), the proton energy-spectra show distinct power-law indices between the low-energy VD part and the high-energy IVD part, implying different acceleration phases or mechanisms, consistent with the previous evidence.

The physical origin of the IVD feature can be explained using the diffusive shock acceleration (DSA) theory. In DSA, particles gain energy by repeated shock crossings while scattering off converging magnetic irregularities; the energy gain and the resulting spectrum depend on the upstream/downstream flows and turbulence (Vainio, 1999). Using the IVD onset spectral index γ and the measured upstream speed $u_u$ in the shock frame, the shock compression ratio $r$ and downstream speed $u_d$ are derived based on previous theoretical work. With the parameters of the shock acceleration conditions retrieved from the DSA model, the energy-dependent acceleration time scale under the shock conditions (Drury, 1983; Prinsloo et al., 2019) is further determined by:

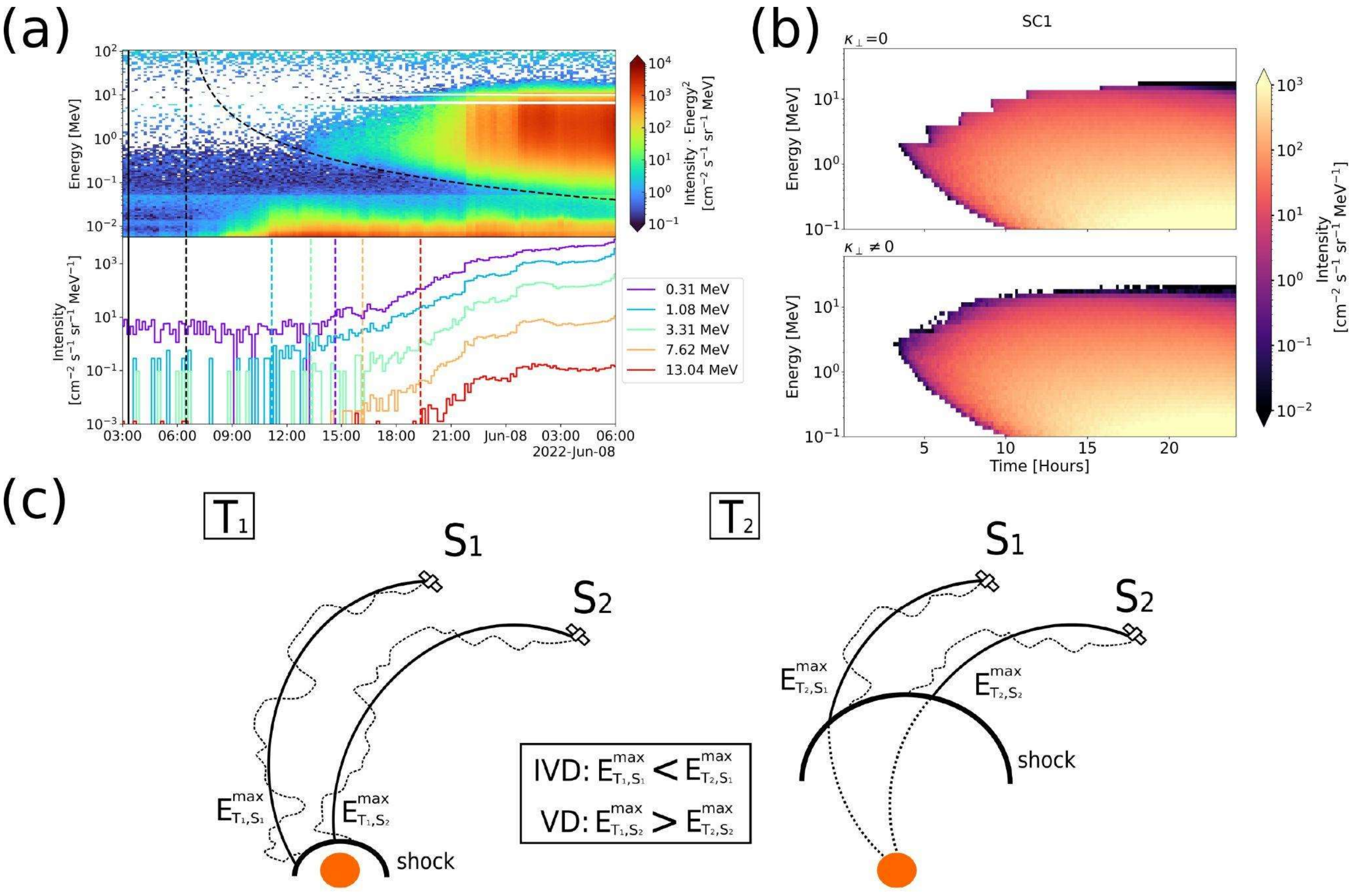

**Figure 18.** Summary of the SEP event observed by SolO on 2022 June 7 and its model results. (a) Proton dynamic spectrum (upper) and time-intensity profiles (lower). The curved dashed line in the upper panel represents the proton onset times fit. The colored vertical lines in the lower panel indicate the onset times for each energy channel. The solid and dashed vertical lines mark the CME eruption and release times, respectively. (b) Modeled proton dynamic spectra observed at SolO. The upper panel shows the case without cross-field diffusion, whereas the lower panel shows the case with cross-field diffusion. (c) Schematic of the shock connectivity at two time steps (see text). Adapted from Ding et al. (2025).

$$\tau_a = \frac{3}{u_u - u_d} \int_{p_{inj}}^{p} \kappa_{rr} \left(\frac{1}{u_u} + \frac{1}{u_d}\right) \frac{dp'}{p'} \qquad (2)$$

Here $\kappa_{rr}$ is the effective diffusion coefficient, expressed by $\kappa_{rr} = \frac{v}{3} \cdot \lambda_{rr} = \frac{v}{3}\lambda_0 (\frac{R}{R_0})^{1/3}$ where $v$ is the particle speed, $R$ is the particle rigidity and $\lambda_0$ is a reference mean free path at $R_0 \equiv 1$ GV. As shown by the resulting trend in Figure 17(c), larger released final energy requires larger acceleration time.

In DSA, the ability of the shock to confine particles for continuous acceleration is parameterized by the particle mean free path $\lambda_0$: smaller values of $\lambda_0$ lead to better confinement and more efficient acceleration of particles. The observationally derived energy-dependent acceleration time is then used to retrieve the theoretical mean free path of particles at the acceleration shock close to the Sun. Fitting the observed IVD release-time (markers in Figure 17(c)) curves for the 2023-11-09 event yields $\lambda_0 \sim 10^{-4}$ au (dashed line in Figure 17(c)), implying strong turbulence and efficient confinement near the acceleration site. $\lambda_0$ inferred by Li et al. (2025) from the other two events are of the same order of magnitude. Although more statistics are needed for a more general conclusion, this is indicative of common characteristics of the DSA process, e.g., $\lambda_0$ is determined by the excitation of Alfvén waves near the shock front by streaming protons (Li et al., 2005).

Li et al. (2025) explain the IVD feature by the longer time required to accelerate higher-energy particles in the DSA mechanism, based on which the shock parameters during the acceleration process are derived. Therefore, this new trait in the high-energy range of solar proton events helps to innovatively retrieve physical parameters during the acceleration process directly from observations, in particular the actual acceleration time scales that cannot be directly observed.

Among the IVD events observed by SolO, some exhibit long durations of IVD, exceeding 10 hours from the nose energy to maximum energy. As the shock propagates outward, the magnetic connectivity between the observer and the shock naturally evolves. In such cases, shock connectivity, the evolution of the shock itself, and transport effects cannot be disregarded. Addressing this complexity, Ding et al. (2025) investigate a pronounced IVD during the 2022 June 7 SEP event. Figure 18(a) shows classical VD is evident below ~ 1 MeV, whereas an IVD extends from ~ 1 MeV to ~ 20 MeV for nearly 10 hours.

Ding et al. (2025) modeled the event using the HEPAT framework (Ding et al. 2022; 2024), which couples the EUHFORIA model (Pomoell and Poedts 2018) to simulate a time-evolving CME-driven shock and the iPATH model (Hu et al. 2017) to simulate energetic particle acceleration and transport in the inner heliosphere. Based on the time-dependent shock properties simulated by EUHFORIA, the local maximum energy $E_{max}$ is calculated by balancing the shock dynamical timescale and the shock acceleration timescale (Zank et al. 2000), which explicitly considers the role of shock acceleration time. Model results show an evolving magnetic connectivity in which the shock speed gradually increases along the field line connecting to SolO out to about 0.4 au. As a result, $E_{max}$ along that field line also increases with time. Notably, the increasing $E_{max}$ here is not primarily a consequence of the increasing acceleration time, but mainly reflects the sampling of different shock locations with increasing shock strength. Figure 18(b) shows the modeled dynamic spectrum is obtained from the transport model. The upper panels represent the case without cross-field diffusion in the transport process. The IVD pattern is evident, with a nose energy around ~ 2 MeV. This behavior directly reflects the evolution of the maximum proton energy along the magnetic field line connecting to SolO. The lower panel displays modeled results that include cross-field diffusion, which enables the observer to sample a broader region of the shock front. It reduces the duration of the IVD phase, since higher-energy particles initially released from different regions of the shock can reach the observer earlier by traversing field lines. This highlights an important point: to observe clear IVD signatures, the influence of cross-field diffusion should be limited. Figure 18(c) sketches shock connectivity at two times, $T_1$ (eruption) and $T_2$ (later). Two observers, $S_1$ (poorly connected) and $S_2$ (well connected), link to the shock via mean (solid) and meandering (dashed) field lines, with the connected maximum proton energy denoted by $E^{Max}_{Tx,Sx}$. For $S_1$, connectivity strengthens with time so that $E^{Max}_{T2} > E^{Max}_{T1}$, producing delayed access to higher energies and an inverse velocity dispersion; for $S_2$, $E^{Max}_{T2} < E^{Max}_{T1}$, yielding classical velocity dispersion. The study demonstrates that long-duration IVD is a natural outcome of evolving magnetic connectivity to a spatially non-uniform CME shock. These results predict an east–west asymmetry: IVD is most likely when the observer first connects to the weaker western flank and later migrates toward the nose.

The combined effects of DSA energy-dependent acceleration and transport have been previously considered in some studies for the 'inverse velocity arrival' event observed by PSP at 0.07 au on September 05, 2022 (see Section 2.2). Kouloumvakos et al. (2025) suggested that it might be attributed to a relatively slow, ongoing particle acceleration process that occurs at the flank of the expanding shock wave. Do et al. (2025) explained the IVD and VD features together by diffusive streaming ahead of the shock of particles that are being accelerated as the shock approaches PSP. In their DSA model, the arrival time at a given energy is the sum of an acceleration delay and a propagation delay for upstream diffusion. When the acceleration delay dominates at higher energies, the relative delay flips from negative to positive, producing the nose pattern. Chen et al. (2025) simulated the same event with the PARMISAN framework. In their picture, the onset delay also includes acceleration time and transport time from the shock to the observer. The key to these interpretations lies in that PSP was initially magnetically connected to a weak flank of the shock, so the acceleration time for high-energy particles was large and became comparable to the transport time along the field line.

Beyond the physical interpretation, observations of SEP events may show what appears to be IVD due to limited detector sensitivity. In general, SEP detectors do not register the actual arrival time of the first particles of an event, but rather the time at which the flux exceeds the detection limit of the specific detector. In the case of gradually increasing fluxes and intensities that decrease sharply with increasing energy, this may lead to the possible observation of IVD.

## 6. Summary

The current era of SEP studies is remarkable and unprecedented in the wealth of observations. Not only are SEP events observed at locations longitudinally surrounding the Sun, they also span 0.05 to > 1 au in distance from the Sun. The energy range coverage and temporal resolution of these measurements are also unparalleled. Given all this, it is not surprising that new, unexpected features of SEP events have been observed. This review highlights a few of them, but certainly not all, with a focus on multi-spacecraft observations of shock-accelerated, large SEP events.

The SEP event of 2022 September 5 was one of the closest to the Sun events ever measured. The fortuitous, near-radial alignment of PSP at 0.07 au and SolO at 0.7 au provided the opportunity to not only examine the evolution of the SEP population, but also of the accelerating shock itself. When PSP observed the SEP event, the shock was still developing and initially incapable (at least on the flank where PSP crossed the shock) of accelerating particles above a few MeV. In contrast, as the shock moved towards SolO, it strengthened sufficiently to produce >90 MeV particles. Detailed analysis of the shock properties at PSP and SolO reveal a shock increasing in complexity as it propagated through the interplanetary medium.

Longitudinally widespread events have been observed in greater detail than ever before and two such events are highlighted in this review. The 2022 February 15 SEP event was observed by more than 8 spacecraft, as well as a Martian rover, spanning 0.35 to 1.5 au and 165° in longitude. Numerous aspects of this event have been examined including unexpected anisotropies at several spacecraft, which indicate multiple sites of particle acceleration both near the Sun and beyond 1 au. This event was one of the few detected 'ground level enhancements' on the surface of Mars. During this event, PSP and BepiColombo were in close proximity providing a unique opportunity to cross calibrate the SEP sensors on the two spacecraft. Such intercalibrations are critical for utilizing the data from the two missions in multi-spacecraft SEP studies. Finally, examination of the particle intensities at PSP as the shock crossed the spacecraft yielded valuable information regarding the diffusion characteristics upstream, near and downstream of the shock at 0.35 au.

Similar to the 2022 February event, the 2022 January 20 SEP event was widespread and also exhibited unusual anisotropies at several spacecraft. In this case it was possible to examine the presence and influence of a previously launched magnetic cloud on the injection and transport of particles. This work underscored the importance of knowing the condition of the IP medium prior to an SEP event and how that may yield unexpected SEP characteristics.

Finally, the PSP energetic particle measurements in the 2022 September event revealed the first observation of an unexpected feature in the energy spectrograms; one for which the SEP community has yet to agree upon a name/term. The characteristics of higher energy particles arriving later than lower energies has been called 'inverse velocity arrival' as well as 'inverse velocity dispersion' and remains incompletely understood. While it is generally accepted that the strength of the shock and time needed to accelerate higher energy particles is a key contributor, the relative roles of evolving magnetic connection to the shock, transport effects, and cross-field diffusion continue to be debated. It is further unclear what governs how long the feature lasts and to what radial distances it may survive. Fortunately, for this and many of the other features found in the events highlighted here (and others), the current multi-spacecraft observations will continue to provide extraordinary data and opportunities to study SEP events, along with the characteristics of the associated shocks and interplanetary context, in ever more detail advancing the understanding of their generation and evolution.

**Acknowledgments**

C.M.S.C., J.G, L.Y.K., J.G.M, D.J.M., and N.A.S. were supported by the Parker Solar Probe mission as part of NASA's Living with a Star (LWS) program (contract NNN06AA01C). ISʘIS data and visualization tools are available to the community at https://spacephysics.princeton.edu/missions-instruments/isois. Solar Orbiter is a space mission of international collaboration between ESA and NASA, operated by ESA. J.R.P. acknowledges the financial support by the Spanish Ministerio de Ciencia, Innovación y Universidades FEDER/MCIU/AEI Project PID2019-104863RB-I00/AEI/ 10.13039/501100011033 and by EU/FEDER and JCCM/INNOCAM, under project SBPLY/24/180225/000108. Solar Orbiter post-launch work at JHU/APL is supported by NASA contract NNN06AA01C, and at the Southwest Research Institute by NASA 80GFSC25CA035. The Solar Orbiter EPD team also acknowledges support from DLR grant 50OT2002 in Germany. L.R.-G. acknowledges support through the European Space Agency (ESA) research fellowship programme. She also acknowledges the financial support by the Spanish Ministerio de Ciencia, Innovación y Universidades FEDER/MCIU/AEI Projects ESP2017-88436-R and PID2019 104863RB-I00/AEI/10.13039/501100011033. O.E.M. and M.K. acknowledge support in the frame of ESA Space Safety Programme's network of space weather service development and pre-operational activities under ESA Contract 4000134036/21/D/MRP.